\documentstyle{mn}
\input{epsf}

\def\ee{$e^+e^-$}
\def\g{$\gamma$}
\def\nh{$N_{\rm H}$}
\def\af{$A_{\rm Fe}$}
\def\ginga{{\it Ginga}}
\def\asca{{\it ASCA}}
\def\xte{{\it RXTE}}

\def\solm{M$_\odot$}
\def\lh{$\ell_{\rm h}$}

\def\ls{$\ell_{\rm s}$}
\def\lhs{$\ell_{\rm h}/\ell_{\rm s}$}
\def\lnh{$\ell_{\rm nth}/\ell_{\rm h}$}
\def\taup{$\tau_{\rm i}$}

\def\O2p{$\Omega/2\pi$}
\def\Gin{$\Gamma_{\rm inj}$}

\def\rin{$r_{\rm in}$}

\def\kTmax{$kT_{\rm max}$}

\def\chisq{$\chi^2$}
\def\xg{X$\gamma$}
\def\El{$E_{\rm line}$}

\def\gmin{$\gamma_{\rm min}$}
\def\gmax{$\gamma_{\rm max}$}

\title[Cyg X-1 in the soft state]
{Radiation mechanisms and geometry of Cygnus X-1 in the soft state}

\author[M. Gierli\'nski et al.]
{Marek~Gierli\'nski$^{1,2}$, Andrzej~A.~Zdziarski$^2$, Juri~Poutanen$^3$,
Paolo~S.~Coppi$^4$,
\newauthor Ken~Ebisawa$^{5}$ and W.~Neil~Johnson$^6$\\
$^1$Astronomical Observatory, Jagiellonian University, Orla 171, 30-244 Cracow,
Poland\\
$^2$N. Copernicus Astronomical Center, Bartycka 18, 00-716 Warsaw, Poland\\
$^3$ Stockholm Observatory, SE-133 36 Saltsj\"obaden, Sweden\\
$^4$ Astronomy Department, Yale University, P.O. Box 208101, New Haven, CT
062520-8101, USA\\
$^5$Code 660.2, Laboratory for High Energy Astrophysics, NASA/Goddard Space
Flight Center, Greenbelt, MD 20771, USA\\(also  at Universities Space Research
Association)\\
$^6$E. O. Hulburt Center for Space Research, Naval Research Laboratory,
Washington, DC 20375, USA\\}

\date{Accepted  Received }
\pagerange{\pageref{firstpage}--\pageref{lastpage}}
\pubyear{1999}

\begin{document}

\topmargin = -0.5cm

\maketitle

\label{firstpage}

\begin{abstract}
We present X-ray/\g-ray spectra of Cyg X-1 observed during the transition from
the hard to the soft state and in the soft state by \asca, \xte\ and {\it
CGRO}/OSSE in 1996 May and June. The spectra consist of a dominant
soft component below $\sim 2$ keV and a power-law-like continuum extending to
at least $\sim 800$ keV. We interpret them as emission from an optically-thick,
cold accretion disc and from an optically-thin, non-thermal corona above the
disc. A fraction $f \sim 0.6$ of total available power is dissipated in the
corona.

We model the soft component by multi-colour blackbody disc emission taking into
account the torque-free inner-boundary condition. If the disc extends down to
the minimum stable orbit, the \asca/\xte\ data yield the most probable black
hole mass of $M_{\rm X} \approx 10$\solm\ and an accretion rate, $\dot{M}
\approx 0.5 L_{\rm E}/c^2$, locating Cyg X-1 in the soft state in the upper
part of the stable, gas-pressure dominated, accretion-disc solution branch.

The spectrum of the corona is well modelled by repeated Compton scattering of
seed photons from the disc off electrons with a hybrid, thermal/non-thermal
distribution. The electron distribution can be characterized by a Maxwellian
with an equilibrium temperature of $kT_{\rm e}\sim 30$--50 keV and a Thomson
optical depth of $\tau \sim 0.3$ and a quasi-power-law tail. The compactness of
the corona is $2 \la \ell_{\rm h} \la 7$, and a presence of a significant
population of electron-positron pairs is ruled out.

We find strong signatures of Compton reflection from a cold and ionized medium,
presumably an accretion disc, with an apparent reflector solid angle, \O2p\
$\sim$ 0.5--0.7. The reflected continuum is accompanied by a broad iron
K$\alpha$ line.
\end{abstract}


\section{Introduction}
\label{sec:introduction}

Cyg X-1 is the best-studied Galactic black-hole candidate. Its X-ray source was
discovered in a 1964 June rocket flight (Bowyer et al.\ 1965). Its optical
companion is HDE 226868 (Webster \& Murdin 1972; Bolton 1972), an OB supergiant
(Walborn 1973). HDE 226868 belongs to the NGC 6871/Cyg OB3 association (Massey,
Johnson \& DeGioia-Eastwood 1995). The distance to the association has  been
measured by Massey et al.\ (1995) to be $D\approx 2.1\pm 0.1$ kpc, whereas
Malysheva (1997) obtained $D\approx 1.8$ kpc using a different photometric
system. Hereafter, we adopt $D =2$ kpc. We note that a commonly-used value of
2.5 kpc was obtained from the lower limit set by the extinction observed
towards Cyg X-1 compared to the extinction--distance dependence for stars with
directions close to that of Cyg X-1 (Bregman et al.\ 1973; Margon, Bowyer \&
Stone 1973). However, this method cannot be used for OB supergiants, in which
case a significant part of the extinction is commonly local (e.g.\ Massey et
al.\ 1995).

The spectroscopic mass function is
\begin{equation}
f_M =\frac{M_{\rm X} \sin^3 i}{(1 + M_{\rm O} / M_{\rm X})^2} = (0.252
\pm 0.010) {\rm M}_{\rm \odot}
\label{eq:mass_function} \end{equation}
(Gies \& Bolton 1982), where $i$ is the inclination of the orbital axis to the
observer, $M_{\rm O}$ and $M_{\rm X}$ are the masses of the supergiant and the
compact object, respectively. The mass ratio, $q\equiv M_{\rm O}/M_{\rm X}$,
has been constrained to be within $1.5\la q \la 2.3$ (Ninkov, Walker \& Yang
1987; Gies \& Bolton 1986). Thus, $M_{\rm X}$ is strongly anticorrelated with
$i$,
\begin{equation}
\label{eq:mrange1}
1.5{\rm M}_\odot \la M_{\rm X} \sin^3 i \la 2.9{\rm M}_{\odot} .
\end{equation}
The companion mass, $M_{\rm O}$, is probably in a range between $\sim 15$\solm\
(Herrero et al.\ 1995) and $\sim 30$\solm\ (Gies \& Bolton 1986). This mass
range together with the above range of $q$ lead to $M_{\rm X}$ between $\sim
6.5$\solm\ and $\sim 20$\solm. Combined with the mass function, these
constraints imply $25\degr\la i \la 50\degr$. This is within limits from
various considerations (e.g.\ Dolan \& Tapia 1989; Ninkov et al.\ 1987; Davis
\& Hartman 1983) yielding $25\degr \leq i \leq 67\degr$ (where the upper limit
is due to the lack of eclipses), and a recent H$\alpha$ Doppler tomography
(Sowers et al. 1998), giving $i\leq 55^\circ$.

The X-ray/\g-ray (hereafter X\g) emission of Cyg X-1 undergoes transitions
between a hard (also called `low') spectral state and a soft (`high') one. Most
of the time the source is in the hard state, where the X\g\ spectrum can be
roughly described by a hard power law with a photon index of $\Gamma
\sim 1.6$--1.8 and a high-energy cut-off above $\sim 100$ keV, and a Compton
reflection component including an Fe K$\alpha$ line (e.g.\ Gierli\'nski et al.\
1997a, hereafter G97). A soft excess observed below $\sim 3$ keV includes a
blackbody with temperature $kT_{\rm bb} \sim 0.15$ keV (Ba{\l}uci\'nska-Church
et al.\ 1995; Ebisawa et al.\ 1996). A likely model for the hard state of Cyg
X-1 embodies Comptonization of blackbody photons from an optically-thick
accretion disc in a hot, thermal, optically-thin plasma.

\begin{figure}
\begin{center}
\leavevmode
\epsfxsize=8.4cm \epsfbox{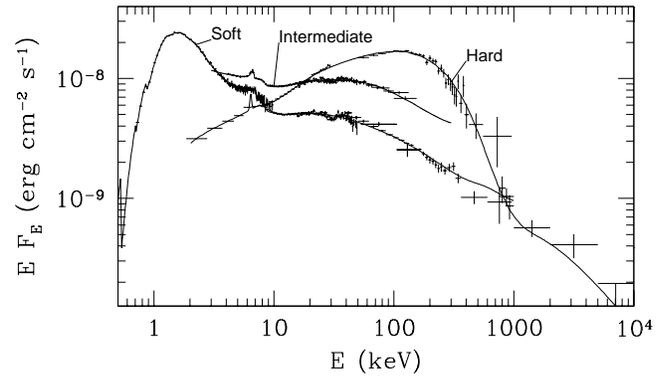}
\end{center}
\caption{Spectral states of Cyg X-1. The hard-state spectrum was observed
by \ginga\/ and OSSE on 1991 June 6 (G97) and by COMPTEL 1991 May 30--June 8
(McConnell et al.\ 1994). The intermediate state was observed by \xte\/ on 1996
May 23 (data set 2 in Table \ref{tab:obslog}). The soft state was observed by
\asca\/ and \xte\/ on 1996 May 30 (data set 3) and by OSSE 1996 June 14--25
(data set 10). Here and in subsequent figures, the spectral data are re-binned
for clarity. The curves correspond to the hybrid model described in Section
\ref{sec:hybrid_model_description}. }
\label{fig:states}
\end{figure}

In the soft state, the X\g\ spectrum is dominated by a soft, blackbody-like
component peaking at $\sim 1$ keV followed by a tail, which can be roughly
described as a power law with $\Gamma\sim 2.5$ (Dolan et al.\ 1977; Ogawara et
al.\ 1982), extending to at least several hundred keV (Phlips et al.\ 1996).
The soft component can be modelled as thermal emission from an optically-thick
accretion disc (Mitsuda et al.\ 1984; Hanawa 1989).  The nature of the hard
tail has remained unclear.  The soft state of Cyg X-1 was observed in 1970,
1975, 1980, 1994 and 1996, each time lasting no more than a few months (Liang
\& Nolan 1984; Cui et al.\ 1997).  In 1994, the soft state was observed only in
hard X-rays and soft $\gamma$-rays by the OSSE and BATSE detectors aboard {\it
Compton Gamma Ray Observatory\/} (Phlips et al.\ 1996; Paciesas et al.\ 1997;
Ling et al.\ 1997), with no accompanying observations in soft X-rays.  Figure
\ref{fig:states} shows  X\g\ spectra of Cyg X-1 in the hard and soft states of
1991 and 1996, respectively, as well as in an intermediate state of 1996 May.

In this paper, we study \xte, \asca\/ and OSSE spectral data from observations
of Cyg X-1 during its soft state in 1996. We find that spectra can be well
described by emission of a non-thermal plasma in the vicinity of an accretion
disc. We study then implications of this model for the size of the plasma and
its location with respect to the disc, the presence of \ee\ pairs, the inner
radius of the accretion disc, the disc stability, and the mass of the central
black hole.


\section{The data}
\label{sec:data}

The soft state of 1996 lasted approximately from May 16 to August 11 (Zhang et
al.\ 1997c). The observation log of \xte\/ and OSSE observations is given in
Table \ref{tab:obslog}.

The \xte\/ data, from the HEASARC public archive, were obtained during
observations with proposal numbers 10412 and 10512. After preliminary
selection, we analysed the data of 1996 May 22, 23, 30, and June 17, 18. During
the reduction, we rejected data affected by the South Atlantic Anomaly and
those for which elevation above Earth's horizon was less then $10^\circ$. We
then
constructed PCA pulse-height spectra summing up counts from all PCA layers and
detectors. We obtained HEXTE spectra separately for cluster A and B. We added a
2 per cent systematic error to each PCA channel to represent residual
uncertainties of calibration. Since dead-time effects of HEXTE clusters are not
yet well established, we allowed free normalization of the HEXTE spectra during
spectral fits. We used the PCA response matrices v3.0 as included in the {\sc
ftools} v4.1 released on 1998 May 8, and the HEXTE response matrices of 1997
March 20.

\begin{figure}
\begin{center}
\leavevmode
\epsfxsize=8.4cm \epsfbox{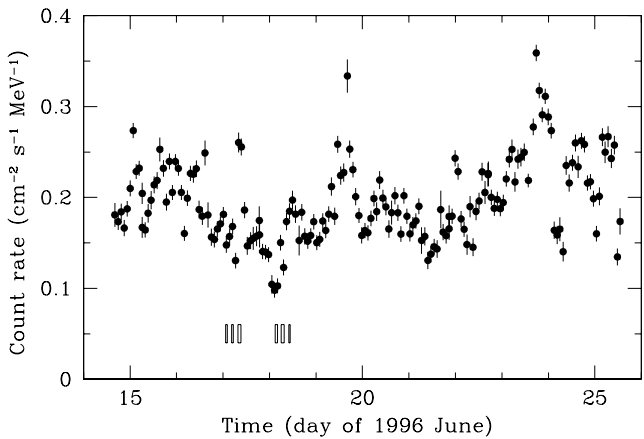}
\end{center}
\caption{The OSSE light-curve in the 50--150 keV band. Each data point
corresponds to one orbit. The boxes in the lower part of the plot indicate the
times of six \xte\ observations (Nos.\ 4--9 in Table \ref{tab:obslog}).}
\label{fig:osse_light}
\end{figure}

OSSE observed Cyg X-1 on 1996 June 14--25 (viewing period 522). The lightcurve 
of this observation is shown in Figure \ref{fig:osse_light}. The flux is 
moderately variable, within a factor of $\sim 2$. The OSSE observation overlaps 
with 6 \xte\ observations on June 17 and 18 (data sets 4--9). We have thus 
extracted 6 OSSE data sets (in the 50--1000 keV range) nearly simultaneous with 
the \xte\ observations.  However, in order to get better statistics, we have 
increased each OSSE data interval by 30 minutes on each side of the 
corresponding \xte\ interval, as shown in Table \ref{tab:obslog}. The OSSE data 
include energy-dependent systematic errors (estimated using both in-orbit and 
pre-launch calibration data), which range from $\sim 3$ per cent at 50 keV to 
$\sim 0.3$ per cent at $\ga 150$ keV. We have also obtained the OSSE spectrum 
from the entire observation (data set 10).  The excellent statistics of this 
spectrum allows a study of the average spectral properties in soft \g-rays, 
e.g.\ possible presence of an annihilation feature or a high-energy cut-off.

\begin{figure}
\begin{center}
\leavevmode
\epsfxsize=8.4cm \epsfbox{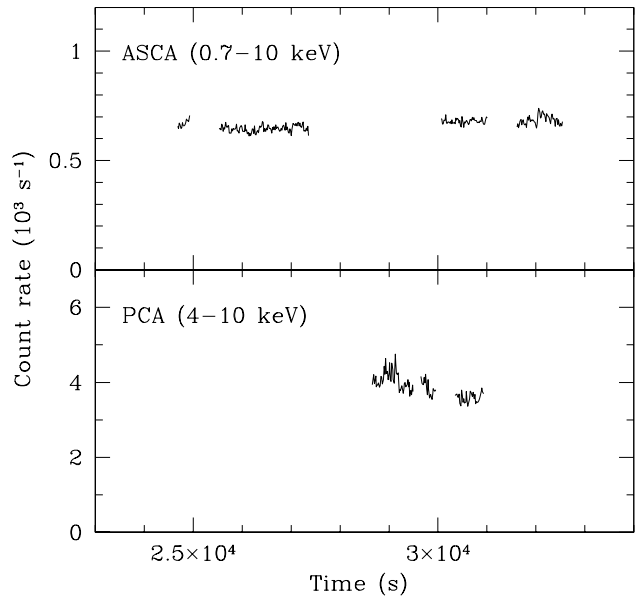}
\end{center}
\caption{The 0.7--10 keV {\it ASCA}/GIS and 4--10 keV {\it RXTE}/PCA
light-curves (time is measured from the beginning of 1996 May 30) of the nearly
simultaneous observations used here (data set 3 in Table \ref{tab:obslog}). }
\label{fig:lightcurve}
\end{figure}

\asca\ observed Cyg X-1 from 1996 May 30 $5^{\rm h}30^{\rm m}$ to May 31
$3^{\rm h}20^{\rm m}$ UT (Dotani et al.\ 1997). The GIS observation was made in
the standard PH mode. The SIS data suffer from heavy photon pile-up and thus
are not usable. The \asca\ and \xte\ observations on May 30 overlap only on a
short period of time. Therefore, in order to improve the statistics, we used
some \asca\ data before and after the \xte\ observation of the data set 3,
which results in 1488 s of the net exposure time (which corresponds to about
4000 s of on-source data). Although the resulting \asca/\xte\ data are not
strictly simultaneous, the constancy of the \asca\ light-curve, shown in Figure
\ref{fig:lightcurve}, strongly suggests constancy of the \asca\ spectrum. We
used the GIS response matrix v4.0.

\begin{table*}
\vbox to110mm{\vfil Landscape table 1 to be inserted here.
\caption{}
\label{tab:obslog}
\vfil}
\end{table*}

\begin{table*}
\vbox to110mm{\vfil Landscape table 2 to be inserted here.
\caption{}
\label{tab:fit_results}
\vfil}
\end{table*}

\begin{figure}
\begin{center}
\leavevmode
\epsfxsize=8.4cm \epsfbox{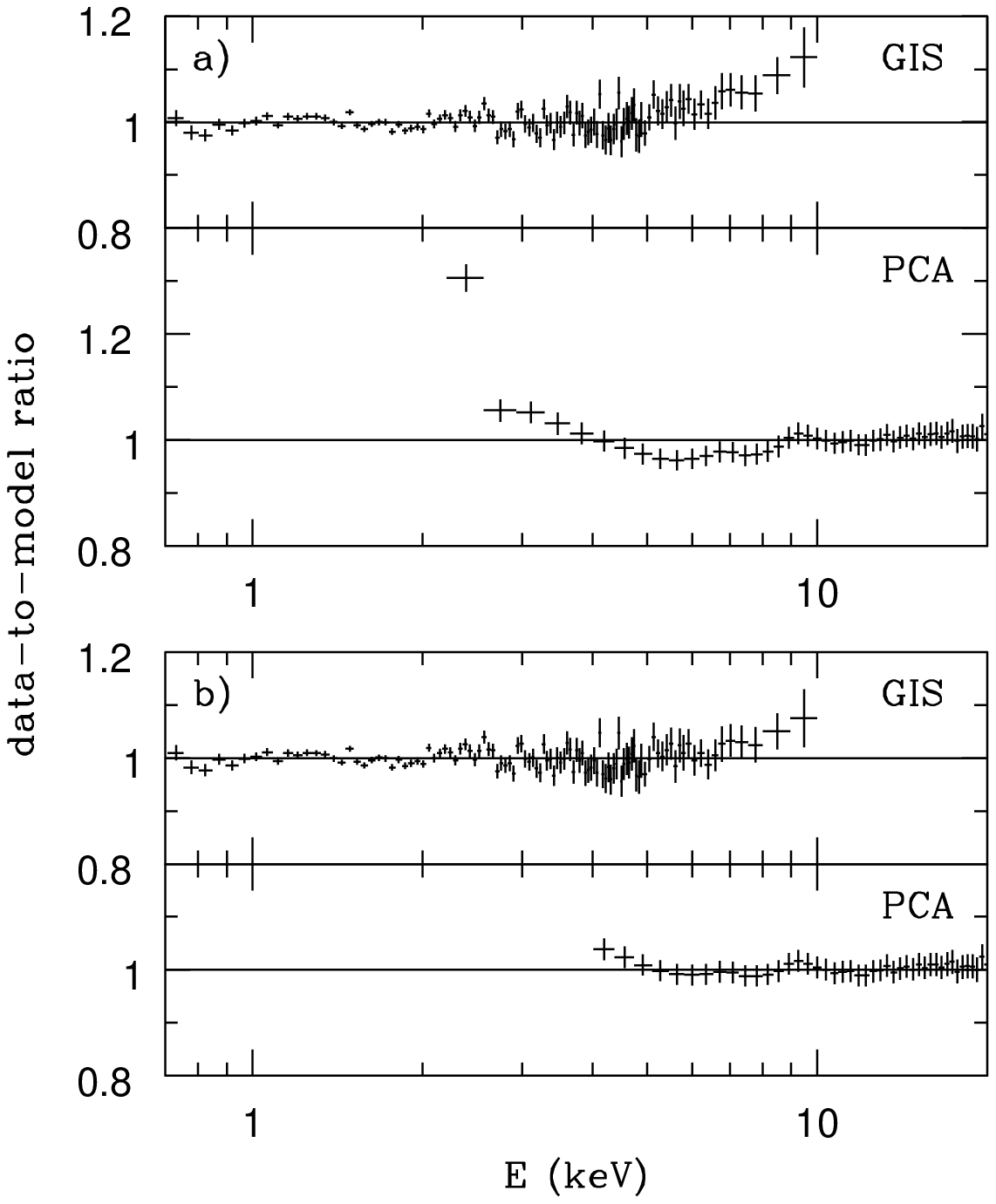}
\end{center}
\caption{The best fit of the non-thermal model to the near-simultaneous \asca\/
and \xte\/ data of May 30 (observation 3). The upper and lower panels show the
data-to-model ratios for \asca/GIS and \xte/PCA, respectively. (a) 0.7--20 keV
residuals of the joint fit to the 0.7--10 keV \asca\/ data and the 2--50 keV
PCA data. (b) The residuals to the fit with the PCA data restricted to the
4--50 keV range. Note a resulting significant improvement of the joint fit.}
\label{fig:asca_pca}
\end{figure}

Near-simultaneous \asca/\xte\ observations of this bright source provide us
with an opportunity to study the cross-calibration of the instruments. Since
the response matrices of X-ray detectors are not diagonal, we cannot compare
detector counts directly. We can only choose a model spectrum, fit it to the
data, and then compare the resulting spectra. For this, we use our best model
(consisting of a disc blackbody and a non-thermal component) found in Section
\ref{sec:hybrid_model_fits} below. We first fit the 0.7--10 keV \asca\ data
simultaneously with the 2--50 keV PCA and, in order to better establish the
high-energy continuum, with the 15--200 keV HEXTE data. Figure
\ref{fig:asca_pca}a shows the data-to-model ratio for the best fit. We see that
there are strong positive residuals at the high and low-energy ends of the GIS
and PCA spectra, respectively, indicating that the PCA spectrum is softer than
the GIS one in the range of the overlap. In particular, in the lowest-energy
range of $\sim 2.2$--2.6 keV, the PCA channel flux exceeds the corresponding
GIS flux by around 30 per cent, which corresponds to about $12 \sigma$. This
discrepancy is probably due to uncertainties in the PCA response matrix,
especially at the lowest energies. A better agreement between the GIS and PCA
data can be obtained if only the PCA channels at $\geq 4$ keV are used, as
shown in Figure \ref{fig:asca_pca}b. Then the PCA-to-GIS flux ratios at 4--4.3
keV and 9--10 keV become 0.96 and 1.08, respectively and we use the 4--50 keV
PCA data hereafter. The above uncertainty in determining the observed X-ray
spectrum will then result in some systematic uncertainty in the parameters of
our theoretical models.

After completion of the data fits presented in this paper, an updated version
of the PCA response matrices was released as a part of {\sc ftools} v4.2 on
1998 December 10. We have then repeated the joint \asca/PCA fit above with the
new PCA response matrix, and found that the discrepancies reported above have
remained virtually unchanged.

\section{Results}
\label{sec:results}

\subsection{Models of absorption, disc emission and reflection}
\label{sec:models}

For spectral fits, we use {\sc xspec} v10 (Arnaud 1996). The confidence range
of each model parameter is given for a 90 per cent confidence interval, i.e.,
$\Delta \chi^2= 2.7$ (e.g.\ Press et al.\ 1992). The spectra are absorbed by
the interstellar medium with a column density \nh, for which we use the
opacities of Morrison \& McCammon (1983) and the abundances of Anders \&
Ebihara (1982). Since only the \asca\ data constrain \nh\ (see Table
\ref{tab:fit_results} below), we assume $N_{\rm H} = 5 \times 10^{21}$
cm$^{-2}$ in all fits not including the \asca\ data.

We fit the observed spectra by a sum of 2 main primary continua. The first one
is due to blackbody emission of an accretion disc. We model it using the
pseudo-Newtonian (hereafter PN) potential and taking into account the
torque-free inner boundary condition, see Appendix \ref{sec:appendix}. The disc
extends from an inner radius, $R_{\rm in}$, corresponding to $r_{\rm in} \equiv
R_{\rm in}/R_{\rm g}$ (where $R_{\rm g} \equiv GM/c^2$) to infinity, and its
temperature distribution is parametrized by the maximum temperature, $T_{\rm
max}$ reached by the disc. We assume $r_{\rm in}= 6$ (i.e., at the minimum
stable orbit in the Schwarzschild metric) unless stated otherwise. We assume $D
= 2$ kpc and, in most fits, the disc inclination of $i=45\degr$ (which is
around the middle of the allowed range of $i$, see Section
\ref{sec:introduction}) and the colour-to-effective temperature ratio, $f_{\rm
col}\equiv T_{\rm col}/T_{\rm eff} = 1.7$ (Shimura \& Takahara 1995).

The second primary continuum corresponds to the excess emission observed above
the high-energy part of the disc emission. For that continuum, we use models
described in Sections
\ref{sec:phenomenological_description}--\ref{sec:hybrid_model_description}.
This continuum is emitted outside the disc, and thus it is Compton-reflected
from the disc surface.

For modelling Compton reflection, we use viewing-angle-dependent Green's
functions of Magdziarz \& Zdziarski (1995), which assume an isotropic point
source (or, equivalently an optically thin corona) above a slab. However, we
treat $\Omega/2\pi$ as a free parameter, where $\Omega$ is an effective (i.e.,
corresponding to the observed strength of reflection) solid angle subtended by
the reflector. The reflection is accompanied by an Fe fluorescence K$\alpha$
line centred at an energy, \El. When the reflection comes from a fast rotating
disc in a strong gravitational potential, Doppler and gravitational shifts
become important. We approximate these effects convolving {\it both\/} the
reflected component and the Fe K$\alpha$ line with the Schwarzschild disc line
profile of Fabian et al.\ (1989). For simplicity, the emissivity of the
reflection and of the line as a function of a disc radius is assumed to vary as
$r^{-s }$. We expect $s \sim 0$, 2, and 3 from different parts of the disc,
depending on the geometry (see Fabian et al.\ 1989). We assume $s = 2$ unless
stated otherwise. In calculating the relativistic distortion, we consider a
range of radii from \rin, which equals that used for the disc blackbody
emission, to $r_{\rm out} = 1000$.

The reflected medium is photo-ionized, and the ionization is described by a
parameter $\xi$. To allow direct comparison with other results, we follow Done
et al.\ (1992) in the definition of the ionization parameter as $\xi \equiv 4
\pi F_{\rm ion}/n$, where $F_{\rm ion}$ is the 5 eV--20 keV irradiating flux in
a power-law spectrum and $n$ is the density of the reflector. (We note that in
the soft state of Cyg X-1, most of the ionizing photons have energies $\ga 1$
keV, and a power-law shape of the irradiating continuum is only a rough
approximation.) For simplicity, we assume that $\xi$ is uniform in the medium.
Since the ionization state of the disc close to the black hole will vary
significantly with radius, the value of $\xi$ found from the fits corresponds
to an average ionization of matter at a range of radii from which most of
reflection originates. The abundances in the reflector are of Anders \& Ebihara
(1982) except the relative Fe abundance, \af, which is allowed to be free in
some models. The ion edge energies and opacities are from Reilman \& Manson
(1979) except that now the Fe K-edge energies taken from Kaastra \& Mewe
(1993).

In our fits, we allow for free relative normalization of data from different
instruments, except for the simultaneous PCA/OSSE data. In that case, we found
their relative normalization is fully consistent with unity.

\subsection{Phenomenological description}
\label{sec:phenomenological_description}

As illustrated on Figure \ref{fig:states}, the \xg\ spectrum of Cyg X-1 in the
soft state consists of a blackbody-like soft component dominant below $\sim 2$
keV and a power-law tail extending to at least several hundred keV. In order to
allow a comparison with previous results (Dotani et al.\ 1997, Zhang et al.\
1997b) as well as to establish a phenomenological description of the spectral
data, we start our analysis with a simple model consisting of a PN disc and a
power law with its corresponding Compton-reflection continuum and an Fe line
computed taking into account the relativistic smearing. First, we fit the
simultaneous \xte/OSSE data of June 17 and 18 (data sets 4--9). The photon
spectral index, $\Gamma$, varies between about 2.3 and 2.5, which confirms the
results of simultaneous ASM and BATSE monitoring of Cyg X-1 (Zhang et al.\
1997b). The reflector solid angle, \O2p, is $\sim 0.6$--0.7, and
$\chi_\nu^2\sim 500/475$.

\begin{figure}
\begin{center}
\leavevmode
\epsfxsize=8.4cm \epsfbox{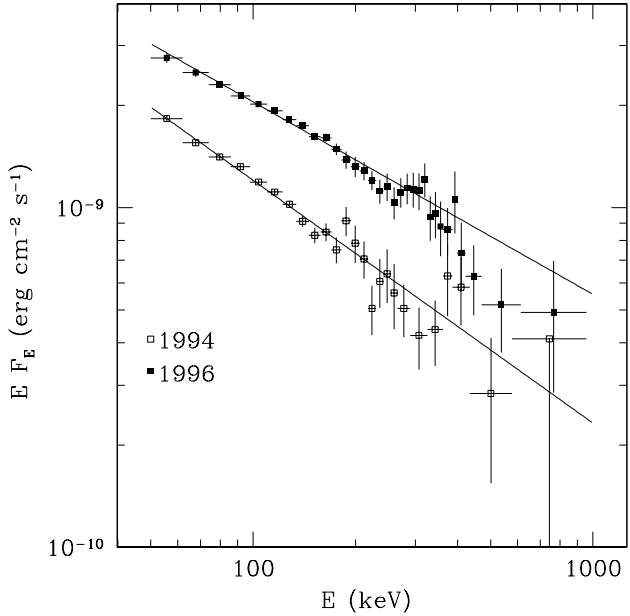}
\end{center}
\caption{OSSE spectra of Cyg X-1 in the soft state of 1994 February (open
boxes) and 1996 June (filled boxes). The former data come from the OSSE viewing
period 318, 1994 February 1--8 (Phlips et al.\ 1996). The latter data are
described in Table \ref{tab:obslog}, observation 10. The two data sets are
fitted by a power law with $\Gamma=2.72^{+0.03}_{-0.04}$, and
$\Gamma=2.57^{+0.02}_{-0.03}$, respectively (solid lines).}
\label{fig:osse}
\end{figure}

Next, we test the \xte/OSSE data for a presence of a high-energy cut-off. We 
multiply the power law by an exponential factor with an e-folding energy, 
$E_{\rm f}$. Since the simultaneous \xte/OSSE observations constrain $E_{\rm 
f}$ only weakly, we test the properties of the high-energy tail using the OSSE 
data integrated over the whole viewing period 522 (data set 10), which spectrum 
is detected up to $\sim 800$ keV (Figure \ref{fig:osse}). When we fit together 
the \xte\ data No.\ 5 and the integrated OSSE spectrum, the best-fit e-folding 
energy is very high, $E_{\rm f} = 940^{+800}_{-310}$ keV.

On the other hand, a power law hard continuum fails to reproduce the low-energy
spectrum of the simultaneous \asca/\xte\/ observation, which was pointed out by
Gierli\'nski et al.\ (1997b) and C98. A fit of a model consisting of a
blackbody disc, a power law and the corresponding reflection continuum with an
Fe K$\alpha$ line provides a very poor fit to this data set, with
$\chi_\nu^2\sim 1100/576$. A much better fit, with $\chi_\nu^2 = 608/574$, can
be obtained with a broken power-law continuum instead of a single power law. We
find the power-law indices, $\Gamma_1 = 2.67_{-0.03}^{+0.02}$, $\Gamma_2 =
2.19\pm 0.02$, the break energy, $E_{\rm break} = (10.9\pm 0.5)$ keV, \O2p\ =
0.48$^{+0.09}_{-0.08}$, \nh\ $= (5.8\pm 0.1) \times 10^{21}$ cm$^{-2}$, \kTmax\
$= 372^{+5}_{-7}$ eV, $M_{\rm X}=9.4^{+0.4}_{-0.2}$\solm, $\xi = (2000\pm 600)$
erg cm s$^{-1}$, $E_{\rm line} = 6.38^{+0.16}_{-0.15}$ keV, and the line
equivalent width, EW $\approx 110$ eV. These results are relatively similar to
those of Dotani et al.\ (1997) and C98.

Furthermore, we stress that if the tail beyond the disc blackbody is due to
Comptonization of the disc blackbody photons (as it is most likely the case in
Cyg X-1), the tail spectrum would have a sharp low-energy cut-off at $\la 2$
keV. A proper description of the tail is then essential for determining the
disc parameters. Using of a continuum without an intrinsic low-energy cut-off
(e.g.\ a power law) can result in an erroneous determination of the disc
properties and \nh.

\subsection{Thermal plasma model}
\label{sec:thermal_model}

As shown in the previous section, the high-energy power law in the soft state
extends to very high energies ($E_{\rm f}\ga 600$ keV). This contrasts the hard
state, where the power law is cut off above $\sim 100$ keV.  If this spectrum
were emitted by Comptonization of soft photons in a thermal plasma, the
electrons would have to have a temperature of at least several hundreds keV.
Then, in order to account for the observed power-law slope, the optical depth
of the electrons would have to be very low, $\la 0.01$. Subsequent scattering
events in a plasma with such very low optical depth would form distinct humps
in the emerging spectrum, which are clearly not seen in the data. We confirm
this using a thermal-Comptonization model of Coppi (1992), assuming that  seed
photons for Compton up-scattering are those emitted by the disc. This model
(together with a disc, reflection and a line) fitted to the \xte/OSSE spectra
(data sets 4--9) yields unacceptable values of $\chi_\nu^2\sim 1000/476$, much
worse than that obtained in section \ref{sec:phenomenological_description} with
a power-law incident continuum.

Furthermore, we find that the thermal-Comptonization model provides a very poor
fit to the \asca/\xte\/ data, as it is unable to reproduce the broken
power-law shape of the tail. This result agrees with that of Cui et
al.\ (1998, hereafter C98) who could fit those data with thermal Comptonization
only assuming that the temperature of seed photons is much higher than that of
the observed disc photons. Thus, we rule out the thermal Comptonization as a
model for the continuum emission in the soft state of Cyg X-1.

\subsection{Hybrid thermal/non-thermal plasma model}
\label{sec:hybrid_model_description}

Since the high-energy tail in the spectrum cannot be due to emission from a 
thermal plasma, we fit the data using a non-thermal model of Coppi (1992) with 
some modifications (see also Coppi \& Blandford 1992; Zdziarski, Coppi \& Lamb 
1990; Zdziarski, Lightman \& Macio{\l}ek-Nied\'zwiecki 1993 and a recent review 
by Coppi 1999). This model embodies Compton scattering, $e^\pm$ pair production 
and annihilation, $p e^\pm$ and $e^\pm e^\pm$ thermal and non-thermal 
bremsstrahlung, and energy exchange between thermal and non-thermal parts of 
the $e^\pm$ distribution via Coulomb scattering. The seed photons for Compton 
scattering are those emitted by the PN disc. Selected $e^-$ are accelerated to 
suprathermal energies and the thermal part of the $e^\pm$ distribution can be 
additionally heated (apart from Compton heating/cooling and Coulomb heating by 
the non-thermal $e^\pm$). [A similar, thermal/non-thermal, model was used, 
e.g., by Holman \& Benka (1992) for solar flares.]

This model assumes a spherical plasma cloud with isotropic and homogeneous
distributions of photons and $e^\pm$, and soft seed photons produced uniformly
within the plasma (Coppi 1992). The properties of the plasma depend on its
compactness, $\ell\equiv {\cal L}\sigma_{\rm T}/({\cal R} m_{\rm e} c^3)$,
where ${\cal L}$ is a power of the source, ${\cal R}$ is the radius
of the sphere and $\sigma_{\rm T}$ is the Thomson cross section. We
specifically use a hard compactness, $\ell_{\rm h}$, which corresponds to the
power supplied to the electrons, and a soft compactness, $\ell_{\rm s}$,
corresponding to the power supplied in the form of soft seed photons. We assume
${\cal R}=10^8$ cm, but this value is important only for bremsstrahlung
emission (which is negligible in our case), and for the Coulomb logarithm (see
Section \ref{sec:hot_plasma} below).

\begin{figure}
\begin{center}\leavevmode\epsfxsize=8.45cm \epsfbox{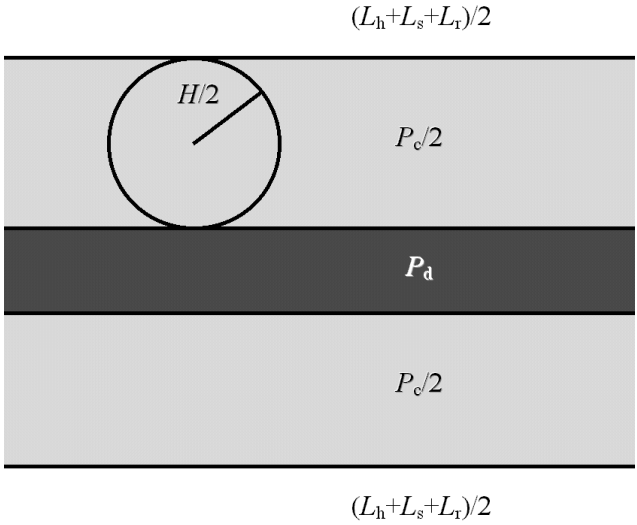}\end{center}
\caption{A sketch of the disc-corona geometry. The emission of the spherical
plasma model is identified with that of the spherical region shown here. Power
dissipated in the disc and corona are $P_{\rm d}$ and $P_{\rm c}$,
respectively. $L_{\rm s}$, $L_{\rm h}$, $L_{\rm r}$ are soft, hard and
reflected luminosities, respectively. See Section
\ref{sec:hybrid_model_description} and Appendix B.}
\label{fig:geometry}
\end{figure}

This spherical model is certainly an idealization of the hot plasma source in
Cyg X-1, and we need to relate the model parameters to those corresponding to
the actual X\g\ source in Cyg X-1. Based on the strength of the blackbody and
reflection components (Section \ref{sec:phenomenological_description}), our
preferred geometry is that of a hot corona above the surface of an accretion
disc. Similar to a method of Bj\"ornsson \& Svensson (1991), we consider a
local spherical volume with a radius ${\cal R}=H/2$ inside the corona, where
$H$ is the corona scale-height, as shown in Figure \ref{fig:geometry}. In some
models, we will also allow the corona to cover the disc at $R\leq R_{\rm h}$
partially with the covering fraction, $g$ ($\leq 1$).

We then consider the relation of the compactnesses, $\ell_{\rm h}$ and $\ell_{\rm
s}$, obtained from fitting the spherical model, to the parameters of the
corona: the luminosity in scattered photons, $L_{\rm h}$, the luminosity in
unscattered blackbody photons, $L_{\rm s}$, the characteristic radius of the
corona, $R_{\rm h}$, and $H$. This will allow us to constrain the last two
quantities by the observed fluxes and fits of the model. For that purpose only,
we assume that the corona radiates isotropically and neglect corrections due
to the finite Thomson optical depth of the corona, the finite disc albedo, $a$,
and the viewing angle of $i\neq 60\degr$ (in which case the anisotropy of the
blackbody flux emitted by the disc plays a role). However, we do take into
account those corrections later in considering the energy balance between the
disc and the corona (Section \ref{sec:balance} and Appendix B).

We point out here one significant difference between the spherical and
disc-corona geometries. Namely, the sphere radiates away {\it all\/} of the
power supplied to the electrons as well as that supplied as seed photons. On
the other hand, the isotropic corona luminosity, $L_{\rm h}$, corresponds to
only {\it half\/} of the total coronal power, $2L_{\rm h}$
(while the other half is radiated towards the disc and absorbed), whereas the
disc luminosity, $L_{\rm s}$, corresponds to all of the power in soft photons
emitted by the disc. In our model fitting, we fit observed fluxes in soft and
hard photons. Therefore, we have to define $\ell_{\rm h}$ in terms of the hard
power radiated away (rather than the dissipated) by electrons in the spherical
volume. This yields
\begin{equation}\label{eq:compactness}
\ell_{\rm h}\approx {1\over 6 g} {L_{\rm h}
\sigma_{\rm T}\over R_{\rm h} m_{\rm e} c^3} {H\over R_{\rm h} }.
\end{equation}
We note that the numerical factor in this equation is not precise due to the
approximations described above, which will introduce some systematic
uncertainty in our determination of plasma parameters.

In models in this Section, we assume $g=1$, which corresponds to all the
blackbody photons passing through the plasma. Then, the soft compactness,
$\ell_{\rm s}$, is to be defined in the same manner as in equation
(\ref{eq:compactness}) except that $L_{\rm h}$ is replaced by $L_{\rm s}$. If
$g<1$, we will distinguish between the soft compactness in photons incident on
the hot plasma and that in soft photons freely escaping (Section
\ref{sec:balance}).

The compactnesses corresponding to the electron acceleration and to the
additional heating of the thermal part of the $e^\pm$ distribution are denoted
as $\ell_{\rm nth}$ and $\ell_{\rm th}$, respectively, and $\ell_{\rm
h}=\ell_{\rm nth}+\ell_{\rm th}$. In our fits, we use the ratio of $\ell_{\rm
nth}/\ell_{\rm h}$ as a parameter. The rate at which non-thermal electrons
appear in the source is assumed to be a power law, $\dot N_{\rm inj}(\gamma)
\propto \gamma^{-\Gamma_{\rm inj}}$, between the Lorentz factors of
$\gamma_{\rm min}$ and $\gamma_{\rm max}$. We hereafter assume \gmin\ = 1.3 and
\gmax\ = 1000. A different choice of \gmax\ has a negligible effect on our
results as long as $\gamma_{\rm max}\gg \gamma_{\rm min}$ since most of the
power in the case of $\Gamma_{\rm inj}>2$ (see Table 2) is injected around
\gmin. The effect of the choice of \gmin\ on our results is discussed in
Section \ref{sec:hot_plasma}.

The injected electrons and pairs produced in photon-photon collisions lose
energy by inverse-Compton scattering (at a rate of $\dot{\gamma}_{\rm Compton}
\propto \gamma^2$), so the equilibrium $e^\pm$ distribution is {\em softer}
then that of injected electrons. This can be seen from the steady-state
continuity equation,
\begin{equation}
{{\rm d} \over {\rm d}\gamma}\left[\dot{\gamma}N_{\rm e}(\gamma)\right] =
\dot{N}_{\rm inj}(\gamma),
\end{equation}
which solution leads to the non-thermal component of $e^\pm$ distribution,
$N_{\rm e} (\gamma) \propto \gamma^{-(\Gamma_{\rm inj}+1)}$. Therefore, the
single scattering of seed photons off these $e^\pm$ forms a power-law spectrum
with photon index $\Gamma \approx \Gamma_{\rm inj}/2 + 1$. In steady state, the
non-thermal and thermal $e^\pm$ have optical depths, $\tau_{\rm nth}$
(typically $\ll 1$) and $\tau_{\rm th}$, respectively, following from balance
of acceleration, pair production, and energy losses. The total optical depth is
denoted by $\tau$.

The optical depth of the thermal component, $\tau_{\rm th}$ (typically $\gg
\tau_{\rm nth}$), is a sum of the optical depth to scattering on $e^+ e^-$
pairs and on $e^-$ coming from ionization of atoms, \taup. The latter one is a
free parameter of the model, but $\tau_{\rm th}$\ as well as the equilibrium
temperature of the thermal component, $T_{\rm e}$, are computed
self-consistently.

The spectrum of the model is then used to compute the corresponding continuum
from Compton reflection including relativistic smearing (Section
\ref{sec:models}). We find that the plasma in the source is so energetic that a
single Compton scattering effectively removes a Compton-reflected photon from
the reflection spectrum. Then, $\Omega/2\pi$ obtained from the fit corresponds
to the unscattered part of Compton reflection.

\begin{figure}
\begin{center}
\leavevmode
\epsfxsize=8.4cm \epsfbox{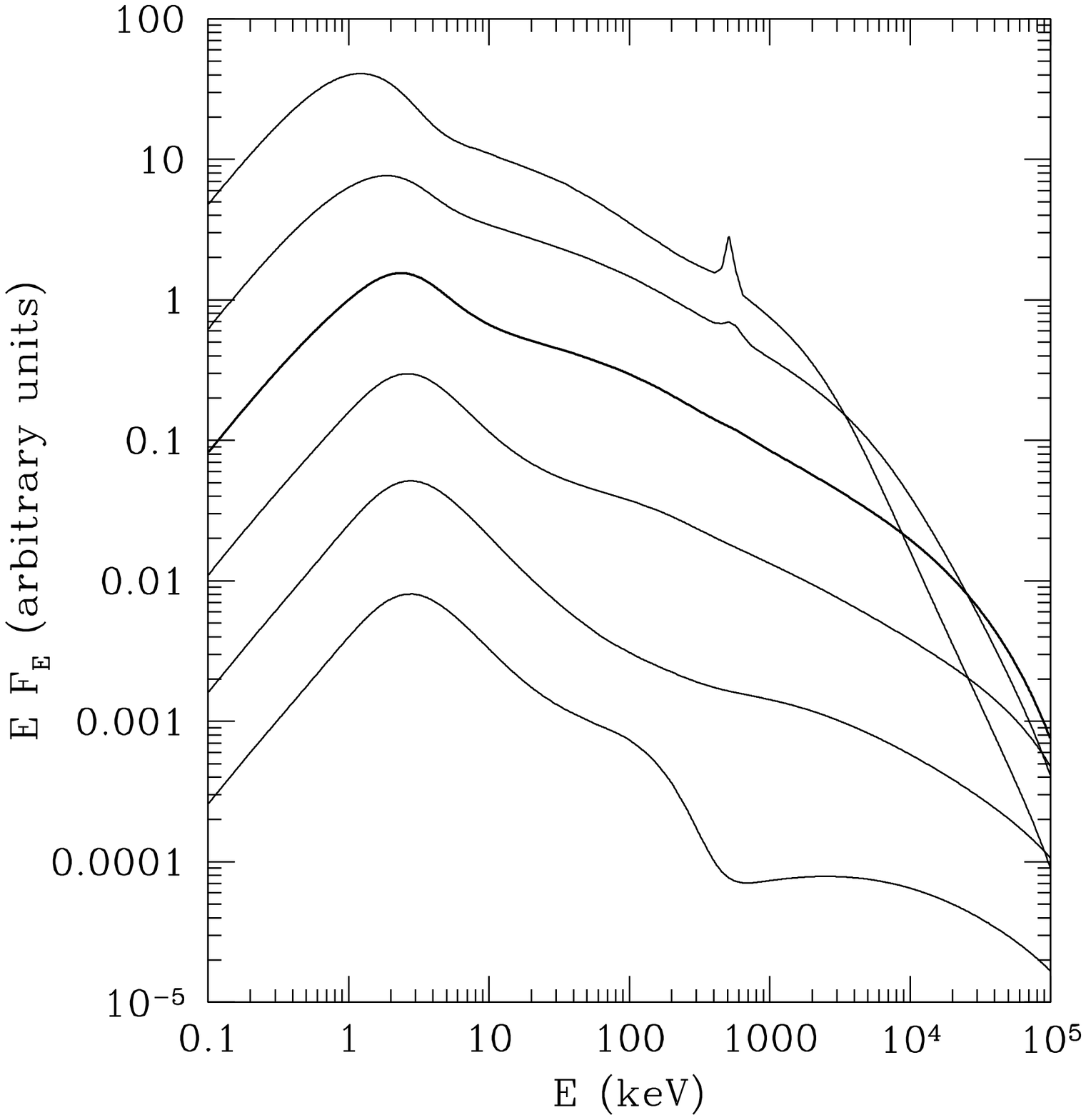}
\end{center}
\caption{Model spectra from non-thermal plasmas above the surface of an
accretion disc calculated for varying soft compactness, \ls. From bottom to
top, \ls\ = 0.01, 0.1, 1, 10, 100 and 1000. Other parameters are \lhs\ = 0.3,
\lnh\ = 0.9, \taup\ = 0.3, \Gin\ = 3, \gmin\ = 1.3, \kTmax\ = 0.4 keV. At low
and high compactness, we see distortions of the hard spectrum due to Coulomb
interactions (see Section \ref{sec:hot_plasma}) and \ee\ pair production and
annihilation (see Section \ref{sec:pairs}), respectively. Compton reflection
and seed photons escaping from the plasma without a scattering are not included
in the spectra shown. The normalization of each spectrum is arbitrary. }
\label{fig:compactness}
\end{figure}

A sequence of model spectra for varying compactness is presented on Figure
\ref{fig:compactness}. For parameters consistent with the observations (as
found in Section \ref{sec:hybrid_model_fits} below) the spectrum is basically a
power law without a break up to $\sim 10^4$ keV. The power law is formed by a
single inverse-Compton scattering of disc seed photons off the non-thermal
$e^\pm$ with optical thickness of $\tau_{\rm nth} \sim 10^{-2}$. At low
energies, we see the spectrum of disc photons repeatedly scattered off the
thermal $e^\pm$, as well as singly-scattered off the nonthermal $e^\pm$. The
former scattering dominates, forming a tail beyond the disc blackbody energies,
crossing the non-thermal power law at $\sim 10$ keV.

\subsection{Spectral fits with the hybrid model}
\label{sec:hybrid_model_fits}

In fitting the data, we have found we cannot uniquely determine the overall
level of the compactness, which, in our fits, is given by the value of
$\ell_{\rm s}$. The joint constraints from both the \asca/\xte\/ and \xte/OSSE
data yield $5 \la \ell_{\rm s} \la 20$. Since the ratio of \lhs\ =
$0.36^{+0.01}_{-0.02}$ is well established (see below), this corresponds to
constraints on the hard compactness, $2 \la \ell_{\rm h} \la 7$. The upper
limit is due to the onset of efficient $e^+ e^-$ pair production at $\ell_{\rm
h}\ga 7$ (see Figure \ref{fig:compactness}), whereas the data do not show any
signatures of the presence of pairs (see Section \ref{sec:pairs}). The lower
limit is due to Coulomb cooling of non-thermal electrons becoming important at
low compactness  (see Figure \ref{fig:compactness}). At small \lh\ this process
thermalizes efficiently the non-thermal electrons below some Lorentz factor
(see Section \ref{sec:hot_plasma}), whereas the data are consistent with a
non-thermal distribution close to a power law. We hereafter assume \ls\ = 10 in
most of fits below, and use $\ell_{\rm h}/\ell_{\rm s}$ rather than $\ell_{\rm
h}$ as a free parameter.

\begin{figure}
\begin{center}
\leavevmode
\epsfxsize=8.4cm \epsfbox{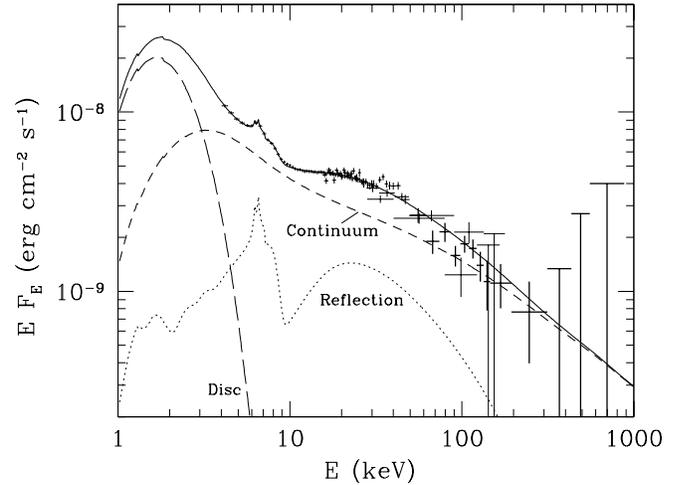}
\end{center}
\caption{The simultaneous \xte\ and OSSE observation of Cyg X-1 in the soft
state on 1996 June 17 (observation No.\ 5) together with a model with
$\tau_{\rm i}=0.25$ (solid curve; fit parameters given in Table
\ref{tab:fit_results}). The model is decomposed into the hot-plasma emission
(short dashes), Compton reflection including the Fe K$\alpha$ line (dots), and
the disc blackbody emission (long dashes). }
\label{fig:juneobs_tau}
\end{figure}

We start spectral fitting with the June 17--18 \xte/OSSE data (observations
4--9). We find that the data do not allow us to determine \taup. There are
several shallow local minima of \chisq\ in a wide range of \taup\ between $\sim
0.2$ and $\sim 3$. However, models with different values of $\tau_{\rm i}$
differ significantly from each other below 4 keV, where we have no data. We
find a strong correlation between \taup\ and \lhs, which latter parameter is
also poorly determined by the data. Therefore, in order to determine \taup\ and
\lhs, we use the \asca/\xte\ observation, for which we obtain
\taup\ $=0.25^{+0.05}_{-0.04}$ and \lhs\ $= 0.36^{+0.01}_{-0.02}$ (see below).
Since the spectral properties did not change significantly between 1996 May 30
and June 17--18, we assume that \taup\ also remained similar, and we fix \taup\
= 0.25 in subsequent fits to the data sets 4--9. Figure \ref{fig:juneobs_tau}
shows the data set 5 fitted with $\tau_{\rm i}= 0.25$. We further note that a
large value of $\tau_{\rm i}$, e.g.\ 2.5, would be difficult to reconcile
within the coronal model, which we find below the most likely, with the
presence of strong Compton reflection in the spectrum.

For the data sets 4--9, we find the best-fitting values of \lnh\ $\sim 0.95$--1
and corresponding 90 per cent confidence intervals covering the range of $\sim
0.8$--1. Thus, in order not to introduce unnecessary free parameters, we assume
\lnh\ = 1 in our fits to the \xte/OSSE data (see Table \ref{tab:fit_results}).

\begin{figure}
\begin{center}
\leavevmode
\epsfxsize=8.4cm \epsfbox{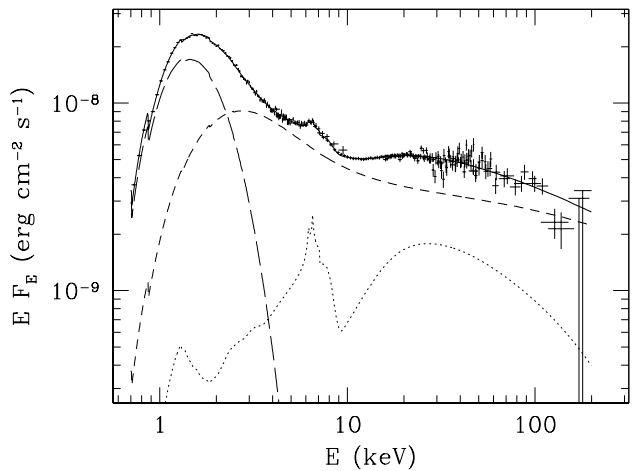}
\end{center}
\caption{The simultaneous \asca\ and \xte\ observation of Cyg X-1 in the soft
state on 1996 May 30 (observation No.\ 3). The short dashes, dots and long
dashes show the Comptonization continuum, the component reflected from the
cold disc and the disc emission, respectively. The solid curve shows the
sum. The fit parameters are given in Table \ref{tab:fit_results}. The
Comptonization continuum is dominated by thermal and non-thermal scattering
below and above, respectively, $\sim 15$ keV, with the resulting spectral break
at that energy. }
\label{fig:mayobs}
\end{figure}

Subsequently, we study the \asca/\xte\ data of May 30. The 0.7--10 keV energy
range of the GIS data allows us to determine the \nh, and therefore we let it
free in this fit. The low-energy coverage also allows us to better constrain
the direct heating of thermal electrons, and thus we let \lnh\ to be free. The
fit results are shown in Table \ref{tab:fit_results}. The best-fitting
parameters correspond to $\tau_{\rm th} \approx 0.26$, $kT_{\rm e} \approx 46$
keV, and $\tau_{\rm nth}\sim 10^{-2}$. We find now the relative non-thermal
power of \lnh\ $= 0.77_{-0.04}^{+0.05}$, which, although somewhat less than
unity, still confirms our conclusion that the direct thermal heating in Cyg X-1
in the soft state is weak compared to non-thermal acceleration. The above range
of \lnh\ is still consistent with the results of June 17--18 \xte/OSSE fits
within the error bars, and thus we are not able to conclude whether the
relative fraction of non-thermal power changed between May 30 and June 17--18
or not. The range of $\ell_{\rm s}$ found above corresponds, at the fitted
$\ell_{\rm h}/\ell_{\rm s}= 0.36^{+0.01}_{-0.02}$, to the total compactness,
$\ell=\ell_{\rm s}+\ell_{\rm h}$, of $7\la \ell \la 27$.

We note here that the actual value of \lnh\ depends somewhat on the assumed
value of \ls. This effect can be understood based on results of Section
\ref{sec:hot_plasma} below. Namely, the importance of Coulomb heating of the
thermal $e^\pm$ plasma component by non-thermal electrons decreases with
increasing compactness. Thus, more direct heating of the thermal $e^\pm$ is
required at a higher compactness than at a lower one. Indeed, for the lower and
upper limit of \ls\ allowable by the data, 5 and 20, we find \lnh\ =
0.88$^{+0.08}_{-0.05}$ and 0.72$^{+0.03}_{-0.02}$, respectively.

On 1996 May 22--23 (data sets 1 and 2), Cyg X-1 was in an intermediate state.
The hard X-ray flux was significantly stronger than in soft-state observations
of May 30 and June 17--18, but weaker than in the hard state (see Figure
\ref{fig:states}). Although Cyg X-1 was observed during this period only by
\xte, we find that the shape of the spectrum allows a determination of \taup.
Fit results are given in Table \ref{tab:fit_results}. Since the soft component
is relatively weak (see Figure \ref{fig:states}), we can only constrain the
seed photon temperature, \kTmax, to be $\la 0.4$ keV, and we thus assume
\kTmax\ = 0.3 keV. We caution here that \kTmax\ is strongly anti-correlated
with $R_{\rm in}$. For example, the data set 1 at \kTmax\ = 0.15 keV yields
$R_{\rm in}$ about 6 times larger than that for $kT_{\rm max}= 0.3$ keV. We
also obtain the compactness ratio of \lhs\ $\sim 1$, which indeed places these
data between the hard and and soft states, for which \lhs\ $\sim 10$ and
0.3--0.4, respectively.

Summarizing our results so far,  we have found a satisfactory physical
description of the data by a model with the main component due to Compton
scattering of soft photons from an optically-thick accretion disc in an
optically thin, hot, plasma. We find the plasma cannot be purely thermal.
Rather, the electrons in the plasma have a hybrid distribution with a
Maxwellian component at a $kT_{\rm e}\sim$ 40--50 keV and a quasi power-law,
non-thermal, tail.

\subsection{$\bmath{e^+e^-}$ pairs}
\label{sec:pairs}

The optical depth of the thermal part of the $e^\pm$ distribution, $\tau_{\rm
th}$, includes contributions from both ionization electrons ($=\tau_{\rm i}$)
and \ee\ pairs. We find first that the \asca/\xte\ observation (data set 3)
cannot constrain the presence of pairs. Fits with $\ell_{\rm s}=40$ and 10 with
a $\chi^2$ difference of only 2.2 yield $\tau_{\rm th}/\tau_{\rm i}\gg 1$ (a
pair-dominated plasma) and 1.07 (an $e^-$ dominated plasma), respectively. This
is because a presence of pairs manifests itself primarily by an annihilation
feature around 511 keV (see Section \ref{sec:hot_plasma}), which energy range
is not covered by \xte.

On the other hand, the \xte/OSSE observations (data sets 4--9) do not constrain
\taup, as discussed in Section \ref{sec:hybrid_model_fits} above, which again
prevents us from constraining $\tau_{\rm th}/\tau_{\rm i}$. Therefore, we fit
the \asca/\xte\ data set 3 together with the average OSSE spectrum (data set
10). The presence of pairs is now excluded at a high statistical significance.
At \ls\ = 10, $\tau_{\rm th}/\tau_{\rm i}\approx 1$, and at \ls\ = 40, which
yields $\tau_{\rm th}/\tau_{\rm i} \approx 1.4$, $\chi^2$ becomes significantly
worse, $\Delta\chi^2 = +16$ at 626 d.o.f. The plasma becomes pair-dominated,
$\tau_{\rm th}/\tau_{\rm i}\gg 1$, at \ls\ = 60, but then $\Delta\chi^2 = +20$.
The main contribution to the increasing $\chi^2$ comes from the OSSE channels
around $m_{\rm e} c^2$ due to an annihilation feature produced by the model but
not present in the data. Thus, pairs appear to be negligible.

In our plasma model, we have assumed so far that the accelerated (or injected)
particles are electrons rather than $e^+ e^-$ pairs (and the acceleration rate
is balanced by the rate at which electrons move from the non-thermal
distribution to the thermal one). On the other hand, some physical
mechanisms may give rise to production of non-thermal pairs rather than electrons, as well
as mostly pairs would be accelerated out of the thermal distribution if that is
already pair-dominated. To investigate that case, we have considered a plasma
model with an injection of $e^+ e^-$ pairs. However, this model gives an
extremely poor fit to the data, $\chi_\nu^2\sim 1000/626$. This is due to the
luminosity from annihilation of the injected pairs, which equals
\begin{equation}
L_{\rm annihilation} \approx {\Gamma_{\rm inj}-2\over (\Gamma_{\rm
inj}-1)\gamma_{\rm min}} L_{\rm h}.
\end{equation}
The annihilation luminosity is then large in our case with $\Gamma_{\rm
inj}\sim 2.5$ (Table 2) and $\gamma_{\rm min}$ constrained to be $\sim 1$ (see
Section \ref{sec:hot_plasma} below), which is in strong disagreement with the
OSSE data showing no pair annihilation feature. Summarizing, the presence of a
substantial number \ee\ pairs in  the X\g\ source of Cyg X-1 in the soft state
is ruled out.

\subsection{Compton reflection and Fe K$\bmath{\alpha}$ line}
\label{sec:reflection}

We find manifest signatures of reprocessing by a cold medium -- an Fe K edge
with a K$\alpha$ line and a broad excess over the continuum at $\sim 10$--100
keV -- in all our data sets. The most straightforward interpretation of those
features is in terms of Compton reflection of the hot-plasma emission by an
ionized, optically-thick, accretion disc. The statistical significance of the
presence of reflection is very high. Fits to the data sets 3--9 without
reflection are much worse, $\chi_\nu^2 \sim 2$, than those including reflection
(see Table \ref{tab:fit_results}), and the main cause of the bad fit are strong
residuals around the Fe edge/line energies. The intermediate-state \xte\ data
sets, 1--2, are somewhat less sensitive to reflection. Still, fits without
reflection are much worse ($\chi^2 = 654$ and 586 at 425 d.o.f.) than those in
Table \ref{tab:fit_results}.

Below, we use the simultaneous \asca/\xte\ observation (data set 3), which
allows the most detailed study of Compton reflection and the associated
K$\alpha$ fluorescence. This is due to a broad energy range, $\sim 0.7$--100
keV, together with good sensitivity near 7 keV. First, we confirm that the
reflected continuum is accompanied by an iron K$\alpha$ line, by means of the
standard $F$-test (e.g.\ Mises 1964). We apply here our best model (see Table
\ref{tab:fit_results}) with relativistic Compton reflection with a line
(described in Section \ref{sec:models}) and without it. Our null hypothesis is
that a model with the line fits the data better than a model without it. A fit
without the line yields $\chi^2_{1} = 652$ with $\nu_1 = 576$ d.o.f., whereas a
fit with the line yields $\chi^2_{2} = 618$, $\nu_2 = 574$ d.o.f. The chance
function, $F_{\rm m} = (\Delta \chi^2 / \Delta \nu) / (\chi_{2}^2 / \nu_2)$ has
a Snedecor's $F$ distribution. The significance level, $P(F > F_{\rm m})$, is
the probability of committing an error of first kind, i.e.\ rejecting the
correct hypothesis. For the fits above, $F_{\rm m} = 15.6$ and $P(F > F_{\rm
m}) = 2.5\times10^{-7}$, which represents a very high statistical significance
that an Fe K$\alpha$ line is indeed present in the data.

We then investigate the issue of the width of the line. For that purpose, we
first compare the \asca/\xte\/ data to a model with static reflection (i.e.,
ignoring the relativistic effects) and a narrow Gaussian line. This yields a
fit worse by $\Delta\chi^2 = +9$ compared to the fit in Table
\ref{tab:fit_results}. This means that indeed the data are much better
described by the line and reflection being relativistically smeared than by
static line and reflection. Still, the disc inner radius is only weakly
constrained by relativistic smearing, to $r_{\rm in}\la 50$ (in which fit we
assumed the inner radius of the blackbody disc emission constant). Also, the
value of the power-law index of reflection emissivity (Section
\ref{sec:models}) is only weakly constrained to $s\sim 2.1^{+0.5}_{-0.7}$. In
order to compare with results of C98, we also considered a broad Gaussian line
in the static-reflection model, obtaining a rather large width of $\sigma_{\rm
Fe} = 1.18^{+0.26}_{-0.22}$ keV (somewhat more than the values given by C98).
Since uncertainties in the {\it RXTE}/PCA response matrix are not yet fully
understood, we have checked our results by ignoring PCA data below 10 keV, in
which case the Fe line parameters were constrained by \asca\ data only. The
results for the line energy and width turn out to be very similar to these
obtained from the overlapping \asca/\xte\ data. Thus, our conclusion is that
the Fe K$\alpha$ line in the soft state of Cyg X-1 is broad.

The reflecting medium is ionized, with $\xi = 350^{+250}_{-120}$ erg cm
s$^{-1}$. Though observations on May 22--23 (data sets 1--2) and June 17--18
(data sets 4--9) yield higher values of $\xi$ (see Table
\ref{tab:fit_results}), those values are relatively weakly constrained and most
of the data are consistent with the \asca/\xte\ result within 90 per cent
confidence. Note that this ionization, obtained with our physical incident
continuum, is much weaker than that found with our phenomenological,
broken-power law, continuum in Section \ref{sec:phenomenological_description},
which shows the sensitivity of the fitted ionization parameter to the shape of
the continuum around 7 keV.

The ionization state found here corresponds to a distribution of Fe ions
peaking at Fe~{\sc xxiii}$^{+{\sc i}}_{-{\sc ii}}$, and with Fe~{\sc
xxiv}--{\sc xxvi} constituting $\sim 30$ per cent of the ion population. Due to
resonant trapping followed by the Auger effect (e.g.\ Ross \& Fabian 1993),
K$\alpha$ photons following ionization of Fe~{\sc xvii}--{\sc xxiii} ions are
completely destroyed and yield no contribution to the observed line. We thus
would expect only a line from Fe~{\sc xxiv}--{\sc xxvi} at $\sim 6.7$--7.0 keV.
Instead, our model yields the rest-frame centre energy of \El\ = $6.37\pm 0.14$
keV for the \asca/\xte\ data and $\sim 6.3$--6.5 keV for other \xte\/
observations. This inconsistency is likely to be due to our simplified model of
reflection, with the ionization state uniform throughout the reflecting medium,
whereas the ionization would, in general, vary considerably with the disc
radius.

We also consider the Fe abundance, \af, of the reflector. In all fits so far,
we have assumed \af\ $= 1$. We find that indeed \af\ = 1 is consistent with the
data. In particular, the \asca/\xte\/ data fitted with \af\ as a free parameter
yield \af\ $= 1.0_{-0.4}^{+0.6}$.

\begin{figure}
\begin{center}
\leavevmode
\epsfxsize=8.4cm \epsfbox{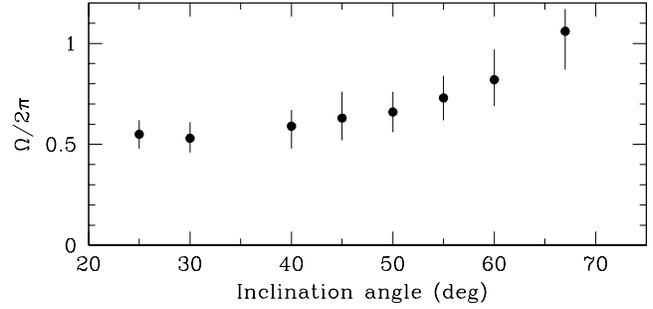}
\end{center}
\caption{The dependence between the solid angle subtended by the reflector,
\O2p, and  the assumed inclination angle, $i$. The error bars are given for 90
per cent confidence.}
\label{fig:reflection}
\end{figure}

One parameter of Compton reflection which remains relatively uncertain is the
inclination angle, $i$. As discussed in Section \ref{sec:introduction}, the
most conservative range of $i$ is $25\degr\leq i \leq 67\degr$. The resulting
$\Omega/2\pi\simeq 0.5$--1.0, increasing with the increasing $i$, is shown in
Figure \ref{fig:reflection}. At the Doppler-tomography limit of $i=55\degr$ of
Sowers et al.\ (1998), $\Omega/2\pi = 0.73\pm 0.11$. Our fits favour large
inclinations angles: $\chi^2 = 626$, 618, 612 (for 574 d.o.f.) at $i =
25^\circ$, $45^\circ$, $67^\circ$, respectively, similarly to a recent result
of Done \& \.Zycki (1999) for the hard state of Cyg X-1. On the other hand, the
plasma and disc properties, e.g., the values of \lhs, \lnh, \taup, $T_{\rm
max}$, are only weakly dependent on $i$.

\subsection{Inclination and the black-hole mass}
\label{sec:disc}

If we assume the disc extends to the minimum stable orbit in Schwarzschild
metric, $r_{\rm in} = 6$, the \asca/\xte\ data yield $M_{\rm X}=
13.3_{-0.8}^{+1.0}$\solm\ at the assumed $D=2$ kpc, $f_{\rm col}=1.7$ and $i=
45\degr$. We note that this mass range is {\it above\/} that allowed by
equation (\ref{eq:mrange1}) at $i=45\degr$. However, as found in Section
\ref{sec:reflection}, the model parameters other than $\Omega/2\pi$ are
insensitive to the assumed value of $i$. Thus, $M_{\rm X}$ can be rescaled to
other values of $i$ using equation (\ref{eq:disc_norm}), which yields
\begin{equation}\label{eq:mrange2} M_{\rm X} = {11.2_{-0.7}^{+0.8} {\rm
M}_\odot \over \cos^{1/2} i }, \end{equation}
which holds for our data to an accuracy better than 2 per cent. This range of
$M_{\rm X}$ overlaps with that of equation (\ref{eq:mrange1}) provided,
\begin{equation}
\label{eq:irange} 0.13 \la {\sin^3 i\over (1-\sin^2 i)^{1/4} }\la 0.28,
\end{equation}
which is satisfied for $29\degr\la i \la 39\degr$, corresponding [equation
(\ref{eq:mrange2})] to $14\ga M_{\rm X}/{\rm M}_\odot\ga 11$. These are the
mass and inclination ranges both consistent with fitting the X-ray data with a
disc extending to the minimum stable orbit and satisfying the constraints on
the mass function and the mass ratio presented in Section
\ref{sec:introduction}.

However, this estimate neglects GR effects in the vicinity of a black hole,
namely gravitational redshift and focusing. We can estimate these effects by
introducing two factors, $f_{\rm GR}$ and $g_{\rm GR}$, correcting the colour
temperature and the integrated flux, respectively (Zhang, Cui \& Chen 1997a;
Cunningham 1975). Then, the correction factor to the mass for a non-rotating
black hole, $r_{\rm in}=6$, and $i \sim 35\degr$ to the mass is $g_{\rm
GR}^{-1/2} f_{\rm GR}^2\approx 0.83$ (with a slow dependence on $i$). Thus, the
above mass range becomes
\begin{equation}\label{eq:mass_gr}
M_{\rm X}\approx (10\pm 1) {\rm M}_\odot,
\end{equation}
which represents our best mass estimate including GR corrections. We note that
this agrees very well with the estimates of Dotani et al.\ (1997) and C98
obtained using a GR disc model (after rescaling from $D=2.5$ kpc used by them
to 2 kpc).

This estimate is sensitive to uncertainties in the distance and the factor
$f_{\rm col}$, with $M_{\rm X} \propto Df_{\rm col}^2$. E.g., a 20 per cent
uncertainty in $Df_{\rm col}^2$ would increase the range of $i$ and $M_{\rm X}$
(after the GR correction) to $27\degr\la i \la 42\degr$, $14\ga M_{\rm X}/{\rm
M}_\odot \ga 7$.

Above, we assumed $r_{\rm in}=6$. However, the data do not require that. When
we allow $r_{\rm in}$ to be a free parameter, the shape of the disc spectrum
changes, and we find this change large enough to constrain \rin\ by the
\asca/\xte\ data. Namely, the model fit to the data is improved by
$\Delta\chi^2 = -6$ at $r_{\rm in}=17^{+26}_{-8}$. Other disc parameters are
now $kT_{\rm max} = 375^{+3}_{-5}$ eV and (at $i=45\degr$ and before the GR
correction) $M_{\rm X} = 10_{-7}^{+10}$\solm. In contrast to the $r_{\rm in}=6$
case, no additional constraints on the disc inclination can now be obtained
from the X-ray data. However, our PN disc model does not take into account
relativistic corrections to the shape of the observed disc-blackbody spectrum.
Thus, this result should be treated simply as a demonstration that the inner
radius of a blackbody disc can, in principle, be constrained by X-ray spectral
fits.

On the other hand, evolution of power spectra, X-ray time lags and the
coherence function indicate that on 1996 May 30 (when the \asca/\xte\/
observation fitted above was performed) the source was still in transition from
the hard to the soft state, and the soft state was established only later, in
particular during the \xte/OSSE observations of 1996 June 17-18 (Cui et al.\
1997). This would then support our finding of $r_{\rm in}>6$ during the
\asca/\xte\/ observation.

\subsection{Energy balance}
\label{sec:balance}

If the hot plasma (either homogeneous or patchy) forms a static corona above
the disc, energy balance can be used to obtain the fraction of the
gravitational energy released in the corona, $f$ (Svensson \& Zdziarski 1994,
hereafter SZ94). The general solution to this problem is given by equation
(\ref{eq:f}) in Appendix B. We assume here isotropic emission of the corona
(corresponding to $d=1$ in that equation). The reflection albedo, $a$, is
computed to be $a\simeq 0.25$ using our \asca/\xte\/ model spectrum at
$i=65\degr$, at which the strength of Compton reflection is approximately equal
to the reflection strength averaged over the $2\pi$ solid angle (Magdziarz \&
Zdziarski 1995). Since the plasma is optically thin, we assume the scattering
probability, $p_{\rm sc}\approx \tau_{\rm i}\approx 0.25$. For the \asca/\xte\/
model spectrum in the case of a homogeneous corona (see Section
\ref{sec:hybrid_model_description}), we find the ratio of the hard flux,
$F_{\rm h}$ (without including reflection) to the soft flux, $F_{\rm s}$, of
$F_{\rm h}/F_{\rm s}\approx 0.50$. (The difference between this ratio and
$\ell_{\rm h}/\ell_{\rm s}$ is due to scattering of soft photons in the
corona.) This corresponds to $L_{\rm h}/L_{\rm s}= 2 (F_{\rm h}/F_{\rm s})\cos
i$ (see Appendix B). At $i=35\degr$ (see Section \ref{sec:disc}), $L_{\rm
h}/L_{\rm s}\approx 0.82$. Equation (\ref{eq:f}) yields then $f\approx 0.63$.
Thus, the geometry with a corona covering most of an inner disc is possible in
the soft state.

On the other hand, the corona may be patchy (Haardt, Maraschi \& Ghisellini
1994; Stern et al.\ 1995; Poutanen \& Svensson 1996). We have investigated that
possibility by adding an additional blackbody disc component to our model. The
emission of that component corresponds to the fraction of the disc photons that
are not passing through the corona. The remaining fraction, $g'$, corresponding
to the disc photons that are incident on the hot plasma, is $g'\geq g$ (where
$g$ the covering fraction defined in Section
\ref{sec:hybrid_model_description}) because the active regions of the patchy
corona can be elevated above the disc surface. We have found that the addition
$g'$ as a free parameter improves the fit by $\Delta \chi^2=-9$, i.e., it is
statistically significant. At the best fit, at which $g'=0.55$, the flux ratio
is $F_{\rm h}/F_{\rm s}= 0.53$. This then gives, from equation (\ref{eq:f}),
$f=0.46$, relatively close to $f$ in the homogeneous-corona case in the
paragraph above.

\subsection{Luminosity and accretion rate}
\label{sec:lum}

The total luminosity in the soft state is, using the \asca/\xte\/ spectrum (see Appendix B),
\begin{equation}\label{eq:l}
L=L_{\rm h}+L_{\rm r}+L_{\rm s}  = 4\pi D^2 \left[ F_{\rm h}(1+a \Omega/2\pi) +{F_{\rm s}\over 2\cos i}\right].
\end{equation}
The observed $F_{\rm h}=3.9\times 10^{-8}$ erg cm$^{-2}$ s$^{-1}$ and $F_{\rm
s}=7.8\times 10^{-8}$ erg cm$^{-2}$ s$^{-1}$ yield, at $i=35\degr$, $L\approx
4.5\times 10^{37}$ erg s$^{-1}$. This corresponds to $\sim 0.03 L_{\rm E}$ at
$M_{\rm X}=10$\solm. Here, $L_{\rm E}$ $\equiv 4 \pi \mu_{\rm e} G M_{\rm X} c
m_{\rm p}/\sigma_{\rm T}$ is the Eddington luminosity, where $\mu_{\rm e} =
2/(1 + X)$ is the mean electron molecular weight and $X \approx 0.7$ is the
hydrogen mass fraction.

If we assume $r_{\rm in}=6$ and no advection, this corresponds to $\dot m\equiv
\dot M c^2/L_{\rm E}\approx 0.5$, or $\dot M\approx 8\times 10^{17}$ g
s$^{-1}$. On the other hand, $\dot M$ can be obtained from fitting the disc
model at known $M_{\rm X}$ and $L_{\rm s}/L$, see equation (\ref{eq:t0}). As
expected, this yields the same $\dot M$ as above.

\section{Discussion}
\label{sec:discussion}

\subsection{Geometry and scaleheight of the corona}
\label{sec:geo}

We consider here constraints on the geometry of the X\g\ source from the plasma
compactness. For the model spectrum for the \asca/\xte\/ data, $L_{\rm h}
\approx 1.9 \times 10^{37}$ erg s$^{-1}$ (see above). The hard compactness is
constrained to $2 \la \ell_{\rm h} \la 7$, which follows from $5 \la \ell_{\rm
s} \la 20$ and \lhs $\approx 0.36$ (Section
\ref{sec:hybrid_model_fits}). From equation (\ref{eq:compactness}), we find the
characteristic radius of the hot corona is then $R_{\rm h}\approx (H/g R_{\rm
h})(1.2$--$4)\times 10^7$ cm. At the black-hole mass of 10\solm\ (see Section
\ref{sec:disc}), this corresponds to $R_{\rm h}\approx (8$--$30)(H/ g R_{\rm
h}) R_{\rm g}$.

We can determine $H$ if we assume hydrostatic equilibrium of the corona. The
pressure in the corona has contributions from radiation, thermal and
non-thermal electrons and ions, and magnetic field. We find that if the ion
temperature, $T_{\rm i}$, equals $T_{\rm e}$, the dominant contribution to
pressure in the corona is from radiation. We use then equation (45) in SZ94 to
estimate $H/R_{\rm h}\sim 10^{-2}$ in the case of a homogeneous corona ($g=1$)
at the radius of maximum disc dissipation per $\log R$, $17R_{\rm g}$.  This
would imply a rather small size of the plasma, $R_{\rm h}\la 0.3 g^{-1} R_{\rm g}$,
violating our assumption of a homogeneous corona, and thus implying $g\ll 1$.
On the other hand, $T_{\rm i}\gg T_{\rm e}$ is a likely condition in many
astrophysical situtations, e.g., two-temperature coronae are considered by Di
Matteo, Blackman \& Fabian (1997). In addition, ions can be accelerated to
non-thermal energies, in addition to accelerated electrons. Then, the corona
may be supported by the ion pressure, together with the pressure of an
equipartition magnetic field.

We can constrain the average energy of ions by requiring that the rate of
Coulomb energy transfer from the ions (both thermal and non-thermal) to
electrons corresponds a luminosity, $L_{\rm Coulomb}^{\rm ie} \ll L_{\rm h}$,
which follows from our fits, in which $\ell_{\rm th}\ll \ell_{\rm h}$. This,
via the condition of hydrostatic equilibirum, constrains in turn the
scaleheight of the corona. We then obtain, using equations (1) and (3) in
Zdziarski (1998),
\begin{equation}\label{eq:h_to_r}
{H\over R_{\rm h}} \approx 2\times 10^2 \Theta_{\rm e}^{3/2} \tau_{\rm th}^{-2} {L_{\rm Coulomb}^{\rm ie} \over L_{\rm E}}
\approx {L_{\rm Coulomb}^{\rm ie} \over L_{\rm h} },
\end{equation}
where $\Theta_{\rm e}\equiv kT_{\rm e} /m_{\rm e} c^2$, and which holds independent of radius. In the second equality, we used the numerical values of
the parameters corresponding to the plasma in Cyg X-1. Then, e.g., $H/R_{\rm
h}=0.3$ (corresponding to $\ell_{\rm th}/\ell_{\rm h}\sim 0.3$, see Section
\ref{sec:hybrid_model_fits}) would correspond to $R_{\rm h}\approx (6$--$20)
R_{\rm g}$ at the covering fraction of $g\approx 0.5$ (Section
\ref{sec:balance}). Thus, the hot plasma can extend over a substantial part of
the disc region where most of the gravitational energy is dissipated. We note
that $H/R_{\rm h}$ of equation (\ref{eq:h_to_r}) corresponds sub-virial ions,
$kT_{\rm i}/m_{\rm p} c^2\sim 0.05/r$.

On the other hand, the corona is not necessarily in the hydrostatic equilibrium
(e.g.\ Beloborodov 1999). Some of the power released in the corona can be used
to accelerate it to some velocity, $\beta c$. We can constrain $\beta$ by
considering the power (or kinetic luminosity), $L_{\rm kin}$, of the outflowing
ions, which equals to the number flux times the bulk-motion energy times area.
This yields
\begin{equation}\label{eq:beta}
{L_{\rm kin}\over L_{\rm h}}\approx {m_{\rm p}\over m_{\rm e}} { \tau
\beta^3\over 2\ell_{\rm h} }.
\end{equation}
For $\tau\sim 0.25$ and $2\la \ell_{\rm h}\la 7$ (Section
\ref{sec:hybrid_model_fits}), $\beta\ga 0.3 (L_{\rm kin}/L_{\rm h})^{1/3}$.
Since no apparent effects of eventual dissipation of $L_{\rm kin}$ are observed
from Cyg X-1, probably $L_{\rm kin}\ll L_{\rm h}$. Then, equation
(\ref{eq:beta}) allows for a mildly relativistic outflow, e.g., $\beta\sim 0.2$
at $L_{\rm kin}\approx 0.2 L_{\rm h}$ and $\ell_{\rm h}=7$.

\subsection{Coulomb interactions and plasma heating}
\label{sec:hot_plasma}

As found in Section \ref{sec:hybrid_model_fits}, the \asca/\xte\/ data require
$7 \la \ell \la 27$. The upper limit is due to effects \ee\ pair production and
annihilation, and the lower limit is due to energy loss of non-thermal
electrons by Coulomb scattering.

The origin of the lower limit has been discussed by Poutanen (1998, see also
Zdziarski et al.\ 1990). For completeness, we discuss it in some detail here.
The energy loss rate due to Compton scattering is (e.g.\ Rybicki \& Lightman
1979)
\begin{equation}
\label{eq:compton_cooling}
\dot{\gamma}_{\rm Compton} \approx  -{4 \over 3} {\sigma_{\rm T} \over m_{\rm
e} c} (\gamma^2 - 1)
U_{\rm ph},
\end{equation}
where $U_{\rm ph}$ is the energy density of photons for which Compton
scattering is in the Thomson limit. For a homogeneous, optically-thin,
spherical source of radius ${\cal R}$ and with luminosity ${\cal L}$, the
energy density is roughly
\begin{equation}
U_{\rm ph} \approx {{\cal L} \over (4/3) \pi {\cal R}^3} {{\cal R} \over c} =
{3 \over 4\pi} {m_{\rm e} c^2 \over \sigma_{\rm T}} {\ell \over {\cal R}},
\end{equation}
where we use $\ell \approx \ell_{\rm s} + \ell_{\rm h}$ for the soft state
of Cyg X-1, in which most of photons in the observed spectrum are in the
Thomson limit. On the other hand, the Coulomb energy loss rate due to
interaction of an electron of $\gamma\gg 1$ and $\gamma- 1\gg \Theta_{\rm e}$ with a background thermal plasma is given by (Gould 1975; Frankel, Hines \& Dewar 1979; Coppi \& Blandford 1992)
\begin{equation}
\label{eq:coulomb_cooling}
\dot{\gamma}_{\rm Coulomb} \approx -{3 \over 2} {c \over {\cal R}} \tau_{\rm
th} \ln(\Lambda\gamma^{1/2} ) 
\end{equation}
where $\Lambda = [2{\cal R}/(3a_{\rm 0}\tau_{\rm th})]^{1/2}$ and $a_0 =5.29
\times 10^{-9}$ cm is the Bohr radius. Then, two rates are equal at a critical Lorentz factor given by
\begin{equation}\label{eq:gamma_cr}
\gamma_{\rm cr}^2 \approx {3\pi \tau_{\rm th}\over 2 \ell} \ln \Lambda + 1.
\end{equation}
Note that $\gamma_{\rm cr}$ depends on ${\cal R}$ only weakly via $\ln
\Lambda$, and thus this results is nearly-independent of the source geometry.
Since $\dot \gamma_{\rm Coulomb}\propto \gamma^0$ while $\dot \gamma_{\rm
Compton}\propto \gamma^2$, the Coulomb energy loss is so fast that it creates a
break at $\gamma_{\rm cr}$ in the non-thermal distribution. This then results
in a break in the distribution of scattered photons, which is {\it not\/} seen
in our data. Also, the energy lost by the non-thermal electrons strongly heats
the thermal plasma, increasing its temperature. These effects are especially
pronounced for $\Gamma_{\rm inj}>2$ (as is the case in our models), when most
of the non-thermal energy is provided to accelerated electrons at the
low-energy end of their distribution.

Since $\gamma_{\rm crit}\propto \ell^{-1/2}$, the observed lack of signatures
of the above effect constrains $\ell$ from below. At the lower limit of $\ell
\approx 7$, and at $\tau_{\rm th} = 0.25$, ${\cal R} = 10^8$cm, $\gamma_{\rm
cr}\approx 2$, i.e., Coulomb interactions just begin to affect the low-energy
end of the non-thermal distribution. Figure \ref{fig:compactness} illustrates
the effect of varying compactness on the spectral shape.

In most our fitted models, we found Coulomb heating of the thermal plasma
sufficient to maintain the plasma temperature required by the observed spectral
shape, $kT_{\rm e}\sim 50$ keV (which, we note, is much higher than the Compton
temperature, $\sim 1$ keV, at which the net energy transfer via Compton
scattering between electrons and photons of the observed spectrum would be
null). Thus, $\ell_{\rm nth}/\ell_{\rm h}\approx 1$ in those models (Section
\ref{sec:hybrid_model_fits}). However, this result depends on \gmin, which has
been set so far to 1.3. A higher value of \gmin\ would decrease the Coulomb
heating and would require some additional heating of the thermal plasma
component. For example, the \asca/\xte\/ data fitted with \gmin\ = 3 gives
\lnh\ = 0.62, i.e., additional heating is required. However, that this fit is
worse by $\Delta \chi^2\approx +9$ with respect to the fit with $\gamma_{\rm
min}=1.3$. Thus, $\gamma_{\rm min}$ is constrained by the data to be $\sim 1$.

Some additional heating of the thermal electrons can be due to the Coulomb
energy transfer from energetic ions, see Section \ref{sec:geo}. The ions can be
thermal or non-thermal. Accelaration of ions is, in fact, quite likely, and not
in conflict with the data. In Section \ref{sec:hybrid_model_fits}, we found
that the electron acceleration rate, $\dot N_{\rm inj}$, is soft, with most of
the accelerated electrons near $\gamma_{\rm min}\sim 1$. If acceleration of
ions follows a similar law, most of accelerated ions will have energies close
to the low-energy end of the power law, and effects of pion and pair production
by ions will be negligible. On the other hand, the accelerated ions will
provide some contribution to Coulomb heating of the thermal electrons (and to
the presure in the plasma), which contribution is allowed by the data as long
as the average ion energy is $\la 50 {\rm MeV}/r$ (see Section \ref{sec:geo}).
We also note that the energetic ions would transfer some energy to the
non-thermal electrons via Coulomb scattering. However, this rate, inversely
proportional to the ion mass, is found by us to be negligible compared to the
energy loss rate of non-thermal electrons to the thermal ones [equation
(\ref{eq:coulomb_cooling})].

Another mechanism of heating the thermal electrons is the so-called
synchrotron boiler (Ghisellini, Guilbert \& Svensson 1988; Ghisellini, Haardt
\& Svensson 1998; Svensson 1999). In this process, thermal electrons are heated
by synchrotron reabsorption of photons emitted by the non-thermal electrons.

\subsection{Anisotropy effects}
\label{sec:anisotropy}

As stated in Section \ref{sec:hybrid_model_description}, our Comptonization
model assumes all the distributions (photons and $e^\pm$) to be homogeneous and
isotropic. This approximation may not hold in some geometries. In particular,
in the disc-corona geometry, the seed photons enter the corona from below only.
Thus, their distribution is anisotropic. Then, a soft photon emitted by the
disc surface is much more likely to scatter on an energetic electron moving
down rather than up. This is because the scattering probability is $\propto
1-(v/c) \cos\vartheta$, where $v$ is the electron speed and $\vartheta$ is the
angle between the directions of the electron and the photon. This anisotropy
affects in different ways spectra from non-thermal and thermal plasmas (and
their $e^\pm$ distributions), and thus we discuss them separately.

The case of a relativistic, non-thermal $e^\pm$ distribution is studied by
Ghisellini et al.\ (1991). Their results show that the angle-averaged downward
flux, due to a single Compton scattering by isotropic electrons with $\gamma\gg
1$, is $\sim 3$ times larger than the flux emitted upwards at the angle of
$45\degr$. This would result in a corresponding strong enhancement of the
Compton-reflected component, which effect is clearly not seen in our data. This
enhancement could be much smaller if the hot plasma forms an inner hot disc and
emission of seed photons and reflection are from an outer optically-thick disc.

On the other hand, the steady-state non-thermal electron distribution in a disc
corona is unlikely to be isotropic. This is because electrons directed downward
lose energy much faster than those directed upward, due to precisely the same
effect as that discussed above. The main mechanism for their isotropisation is
Coulomb interaction since higher-order Compton scatterings are negligible in
our case with $\tau_{\rm nth} \sim 10^{-2}$. The process of Coulomb
interactions of non-thermal electrons with the background thermal plasma is
considered in Section \ref{sec:hot_plasma} above, where we find that the
observed spectrum constrains compactness to a range where the electron energy
losses due to this process are negligible compared to those due to Compton
scattering. Then the negligible energy losses, in the case of $e^\pm e^\pm$
scattering, imply no significant changes of the direction of the relativistic
electrons.  (The rate of interactions of relativistic electrons with protons,
which could change the electron direction without significant energy loss, is
smaller than that with $e^\pm$ by a factor $m_{\rm e}/m_{\rm p}$ and thus even
less significant.) Thus, the steady-state distribution of electrons with
$\gamma>\gamma_{\rm cr}$ [$\ga 2$, equation (\ref{eq:gamma_cr})] will be
strongly anisotropic due to the Compton energy loss rate depending on the
direction.

When this is the case, an isotropic acceleration/injection of relativistic
electron will lead to {\it isotropic\/} emission of singly-scattered photons.
This is because the effects of anisotropic emission and anisotropic energy loss
cancel each other, and the rate of emission in a given direction is simply
proportional to the injection rate (see Ghisellini et al.\ 1991).

Note that an anisotropic electron distribution corresponds to a current in the
plasma. This current will be balanced by an opposite current in the thermal
component of the electron distribution. Since $\tau_{\rm th}\gg \tau_{\rm
nth}$, the latter current will have a negligible effect on the properties of
the thermal plasma component.

Anisotropy effects in the case of a thermal $e^\pm$ distribution are considered 
by Haardt (1993), Stern et al.\ (1995) and Poutanen \& Svensson (1996). They 
are due to the same physical effect as described above, and thus they suppress 
a part of the emitted spectrum due to the first (and to some degree, second) 
order of Compton scattering. On the other hand, photons become closely 
isotropic after a higher number of scatterings. This leads to the so-called 
anisotropy break close to the peak energy of the second-order scattering. In 
contrast to the non-thermal case, we have found that the Coulomb relaxation 
time for the thermal electrons interacting with themselves is about equal to 
the thermal plasma cooling time ($\sim 10^{-4}$ s), and thus no significant 
anisotropy of the thermal electron distribution is expected in our case.

In order to assess the significance of the thermal anisotropy break in the case
of Cyg X-1, we have performed Monte-Carlo simulations for a slab geometry
assuming the plasma parameters as found from data fitting. We have found the
anisotropy break at $\sim 15$ keV, with the outward flux at 7 keV suppressed by
a factor of $\sim 0.8$ with respect to the isotropic case. As seen in Figure
\ref{fig:mayobs} (the dashed curve), this break energy actually coincides with
the break occuring due to intersections of the power laws due to thermal and
non-thermal scattering. Then, plasma parameters slightly different from those
determined neglecting anisotropy effects could still produce the observed
spectrum. Thus, the present data are still consistent with the presence of a
thermal anisotropy break.

On the other hand, if the mildly relativistic outflow is present in the corona
(see Section \ref{sec:geo}), the seed soft radiation emitted by the disc
is more isotropic in the rest frame of the outflowing plasma, thus reducing the
amplitude of the anisotropy break (see Beloborodov 1998, 1999).

\subsection{Disc stability}
\label{sec:disc_parameters}

We examine here the stability of the optically-thick disc, taking into account
the fraction, $f$, of the gravitational energy being dissipated in the corona
(SZ94). A possible mechanism for energy transfer from the disc to the corona
are magnetic flares (e.g.\ Galeev, Rosner \& Vaiana 1979; Beloborodov 1999). A
part of the disc becomes radiation-pressure dominate and thus unstable (both
thermally and viscously) above a critical accretion rate, $\dot{m}_{\rm cr}$.
We find that a radiation-pressure dominated zone in a PN disc appears first
(with increasing $\dot m$) around $R=15.1 R_{\rm g}$. We then extend results of
SZ94 (who considered the Newtonian potential) to the case of the PN potential,
and derive $\dot{m}_{\rm cr}$ in the presence of external dissipation as,
\begin{equation}\label{eq:m_cr}
\dot{m}_{\rm cr} = 0.64 (\alpha M_{\rm X}/{\rm M}_{\odot}
)^{-1/8}[(1-f)\zeta]^{-9/8},
\end{equation}
where $\alpha$ is the viscosity parameter and $\zeta$ is a parameter in the
vertically-averaged radiative transfer equation, which is set by various
authors between 0.5 and 2. We follow here Shakura \& Sunyaev (1973) and set
$\zeta=1$ (which value is in approximate agreement with calculations taking
into account the vertical disc structure, A. R\'o\.za\'nska, private
communication). We note that relativistic effects, as approximated by the PN
potential, have a stabilizing effect on the disc, i.e., they increase the value
of $\dot m_{\rm cr}$. Specifically, the numerical coefficient in equation
(\ref{eq:m_cr}) is only 0.37 in the Newtonian case (SZ94).

\begin{figure}
\begin{center}
\leavevmode
\epsfxsize=8.4cm \epsfbox{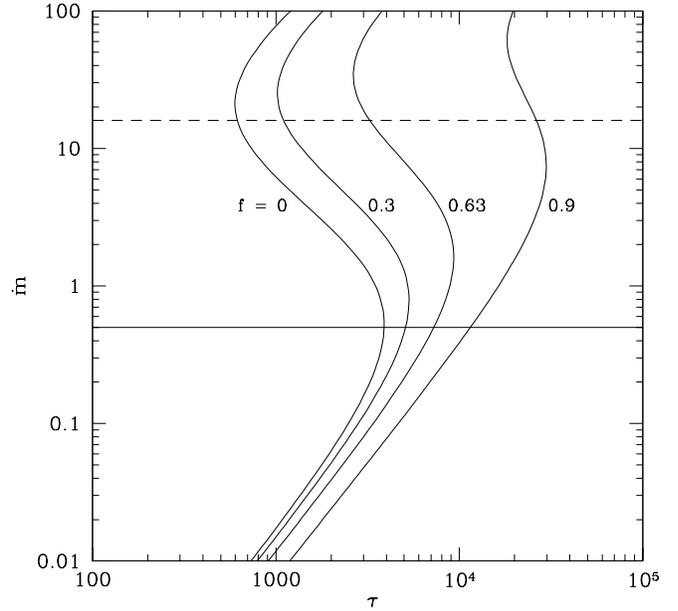}
\end{center}
\caption{The solutions of the optically thick disc (Abramowicz et al.\ 1988)
with a fraction of accretion power, $f$, dissipated in the corona (SZ94),
for $M_{\rm X}=10$\solm\ and $\alpha$ =0.1 at the radius of
the most significant instability of $R = 15.1R_{\rm g}$. The curves are labeled
by their value of $f$, with $f = 0.63$ corresponding for Cyg X-1. The solid and
dashed horizontal lines correspond to the accretion rate of Cyg X-1 in the soft
state and to the Eddington luminosity at an efficiency of $\eta=1/16$,
respectively.}
\label{fig:mdottau}
\end{figure}

At $M_{\rm X}=10$\solm, $f=0.63$, $\dot m_{\rm cr}\approx 1.47 \alpha^{-1/8}$.
This implies that the disc in the soft state of Cyg X-1, with $\dot m\approx
0.5$ (Section \ref{sec:lum}), is stable at any $\alpha$. Figure
\ref{fig:mdottau} shows the solutions of the optically thick accretion disc
with the corona for 4 values of $f$.

On the other hand, the data do not require $r_{\rm in}=6$, and a fit in Section
\ref{sec:disc} gives $r_{\rm in}=17^{+26}_{-8}$. This solution corresponds to a
higher $\dot m$ and a lower $M_{\rm X}$ than in the case of $r_{\rm in}=6$.
Then, $\dot m$ may possibly exceed $\dot m_{\rm cr}$ and the disc instability
may be responsible for the large $r_{\rm in}$.The precise value of $\dot m$ in
this case depends on the unknown partition of the observed luminosity into the
fractions from the corona above the disc and an inner hot flow at $r<r_{\rm
in}$, possibly advective (Narayan \& Yi 1995; Abramowicz et al.\ 1995).
Therefore, the current data do not allow us to distinguish between the
possibilities of $r_{\rm in}=6$ and $>6$.

\subsection{Alternative models}

One alternative model to the non-thermal Compton model presented here is 
thermal Comptonization, which was proposed to model spectra of the soft state 
of Cyg X-1 (Poutanen et al.\ 1997; C98; Esin et al.\ 1998) and other black-hole 
binaries (Miyamoto et al.\ 1991; Esin et al.\ 1997) However, as discussed in 
Section \ref{sec:thermal_model}, this process is unable to account for the 
observed steep power-law tails extending to high energies.

Another possibility for the production of the power-law spectra observed in the
soft state is that of bulk-motion Comptonization (see, e.g., Payne \& Blandford 
1981; Colpi 1988; Laurent \& Titarchuk 1999). If there is quasi-spherical
accretion near the black hole, then cold infalling electrons could acquire
substantial velocities $v \sim c$, and Comptonization using the large bulk
inflow velocity of the electrons (as opposed to their assumed smaller thermal
motions) could give rise to power law spectra like those observed in the soft
state. This model predicts a sharp cutoff in the spectrum at a few hundred keV.
The OSSE data, on the other hand, clearly show no cutoff until at least 800 keV
which strongly rules out  the bulk Comptonization hypothesis.

Also, only a fraction, $\ll 1$, of the disc blackbody photons (emitted 
semi-isotropically with maximum emission per $\log r$ at $r\sim 20$) will be 
incident on the inner spherical inflow (see Figures 1--2 in Laurent \& 
Titarchuk 1999). This appears to lead to a normalization of the tail spectrum 
with respect to the disc blackbody much below the observed normalization. 
Furthermore, the presence of the Compton reflection with a covering factor of 
$\Omega/ 2\pi \sim 0.7$ puts strong constraints on the geometry of the 
accretion flow. Such a covering factor is natural in a disc-corona geometry, 
but it appears incompatible with the bulk Comptonization scenario (Figures 
1--2 in Laurent \& Titarchuk 1999).

\subsection{The nature of state transition}
\label{sec:spectral_state}

A remarkable feature of state transitions in Cyg X-1 is that the associated
change in the bolometric luminosity is very small. We have found $L\approx
4.5\times 10^{37}$ erg s$^{-1}$ in the soft state (Section \ref{sec:lum}),
whereas the maximum bolometric luminosity observed in the hard state is $\sim
3\times 10^{37}$ erg s$^{-1}$ (G97; at $D=2$ kpc). If there is no advection,
the corresponding change of $\dot M$ is small. In the simplest scenario, we
expect no advection in the soft state provided the optically-thick disc extends
to the minimum stable orbit. On the other hand, a commonly used model for the
hard state comprises a hot inner disc (Poutanen, Krolik \& Ryde 1997; Poutanen
\& Coppi 1998; Poutanen 1998; Esin et al.\ 1998), in which case advection is
likely. If the transition from the hard state to the soft one is associated
with an increase of $\dot M$, both that increase and the advected fraction of
the power in the hard state have to be very small (Esin et al.\ 1998). On the
other hand, advection may dominate over cooling in the hard state, in which
case the hard-to-soft transition would be associated with a {\it decrease\/} of
$\dot M$.

In either of the above cases, the hard-to-soft state transition appears to be
associated with a decrease of the minimum radius of the optically-thick disc,
$R_{\rm in}$. When $R_{\rm in}$ is large, the inner disc is hot and it receives
a major fraction ($\sim 0.9$) of the available gravitational energy. The
spectrum is hard, $\Gamma \sim 1.6$, and the covering angle of the reflector is
relatively small, \O2p\ $\sim 0.3$. At a certain moment, the cold disc starts
to move inwards, taking in more and more accretion power, while the hot phase
diminishes. In the final configuration the cold disc extends down to near the
last stable orbit, while the hot (non-thermal) plasma probably forms a corona
above that disc, receiving less of accretion power ($\sim 0.6$) than in the
hard state. This results in a softening of the spectrum, to $\Gamma\sim 2.5$,
and an increase of the strength of the Compton-reflection component, to \O2p\
$\sim 0.7$.

Another possible model for the hard state has recently been proposed by
Beloborodov (1999). In that model, the X\g\ source forms a corona above
optically-thick disc extending close to the minimum stable orbit even in the
hard state. The hardness of the spectrum and the weakness of Compton reflection
in that state are then explained by a mildly-relativistic bulk motion away from
the disc. The hard-to-soft state transition would then involve the corona
becoming static. Possibly, this may be due to the coronal plasma consisting of
\ee\ pairs in the hard state (which is allowed by the form of hard-state
spectrum, G97; Poutanen \& Coppi 1998), and of an ordinary plasma in the soft
state. The data studied here appear not to allow distinguishing between these
possibilities.

\section{Conclusions}
\label{sec:conclusions}

In this work, we have obtained a self-consistent description of the soft state
of Cyg X-1 (observed in 1996 by \asca, \xte\/ and OSSE) in terms of a
(predominantly) non-thermal corona above the surface of an optically-thick
accretion disc. Emission of the non-thermal corona accounts for both the
high-energy power-law tail observed up to several hundred keV and the soft
excess, resolving two problems of the thermal-plasma model of C98.

The non-thermal corona receives most of the supplied power in the form of
electrons accelerated (or injected) to relativistic energies. The acceleration
rate is rather soft, with the power-law index of $\Gamma_{\rm inj} \sim 2.5$--3
above $\gamma_{\rm min}\sim 1$. The electrons lose their initial energy in
Compton and Coulomb processes, and finally form a thermal component (at
$kT_{\rm e}\sim 50$ keV) of their distribution at low energies. The resulting
X\g\ spectrum consists of blackbody photons emitted by the disc (at low
energies) and a component due to Compton upscattering of the disc photons by
both thermal and non-thermal electrons in the corona.

Production of $e^+e^-$ pairs is negligible in the corona, which follows from
the observed lack of an annihilation feature in the spectrum from OSSE. This
constrains from above the compactness (a ratio of the luminosity to the size),
$\ell$, of the source. On the other hand, Coulomb scattering is an important
process in heating the thermal electrons by the non-thermal ones. Still, the
Coulomb energy loss by non-thermal electrons is constrained by the observed
spectrum to be slower than their Compton energy loss, which constrains $\ell$
from below. The resulting constraints are $7\la \ell\la 30$.

We also find that a modest heating of the thermal electrons (in addition to
their Coulomb heating by the non-thermal electrons) is required by the observed
spectrum. This heating can be provided by Coulomb interactions with energetic
ions in the corona, which then also provide most of the pressure support of the
corona.

The coronal geometry of the source is supported by the observed spectral
component from Compton reflection from the disc surface and the Fe broad
K$\alpha$ fluorescence line. Both components are relativistically smeared,
supporting reflection from an inner part of the disc. The strength of the
observed Compton reflection spectrum is fully consistent with reflection from a
slab, after taking into account Compton scattering of reflected photons in the
coronal plasma. The observed softness of the spectrum is consistent with energy
balance in the disc-corona system with $f\sim 0.6$ of the available
gravitational energy dissipated in the corona (and the remainder within the
disc). From the observed luminosity and fitted compactness, we find the corona
covers a substantial fraction of the disc provided most of its pressure support
is provided from energetic ions. Still, the ion energy in random velocities is
constrained to be much below the virial energy (by the constraint on the
Coulomb heating of thermal electrons, see above). Then, the corona can be
static. On the other hand, we find that a mildly relativistic outflow is also
consistent with the observations.

Fits of our PN disc model together with the mass function and constraints on 
the mass ratio imply $M_{\rm X}\simeq (10\pm 1)$\solm\ and the inclination of 
$i\sim 30\degr$--$40\degr$, at the most likely distance of 2 kpc and the colour 
correction of $f_{\rm col}=1.7$.  The bolometric luminosity is $\simeq 
4.5\times 10^{37}$ erg s$^{-1}\simeq 0.03 L_{\rm E}$, which corresponds, in the 
absence of advection, to $\dot m\simeq 0.5$. At that accretion rate, the disc 
is fully stable all the way down to the minimum stable orbit against the 
instabilities associated with the dominance of radiation pressure. On the other 
hand, solutions with advection, a higher $\dot m$ and the disc terminating at 
$r_{\rm in}\gg 6$ are also allowed by the data.

\section*{Acknowledgements}

The authors are grateful to Chris Done, the referee, for valuable remarks, to 
Keith Jahoda for help with the PCA data reduction, to Joanna Miko{\l}ajewska 
for discussions on the distance and binary parameters of Cyg X-1, and to 
Bo\.zena Czerny for providing us with her optically-thick disc code. This 
research has been supported in part by the Polish KBN grants 2P03D00713, 
2P03D00614, 2P03C00511p0(1,4), 2P03D00514, NASA grants and contracts,  the  
Swedish Natural  Science  Research  Council and the Anna-Greta and Holger  
Crafoord Fund, and it has made use of data obtained through the HEASARC online 
service provided by NASA/GSFC.


\appendix

\section{Spectrum from a pseudo-Newtonian disc with zero-stress inner boundary}
\label{sec:appendix}

Let us consider a geometrically-thin, optically-thick, accretion disc around a
Schwarzschild black hole with a mass, $M$. We assume the disk locally emits a
blackbody spectrum, which fraction, $p_{\rm sc}$ is then partly scattered in a
corona. The unscattered flux observed from the disc is given by
\begin{equation}\label{eq:fnu}
F_\nu = (1-p_{\rm sc})  {{2 \pi\cos i} \over {f_{\rm col}^4 D^2}}
\int^\infty_{R_{\rm in}} B_\nu(T) R {\rm d}R,
\end{equation}
where $B_\nu$ is the blackbody spectrum. Other symbols used in this Appendix
are defined in Sections \ref{sec:introduction} and \ref{sec:models} except that
we now drop subscripts in symbols for the colour temperature and mass. We allow
here for truncation of the disc at $R_{\rm in}\geq R_{\rm ms} =6GM/c^2$.

In order to compute the observed spectrum, we need to find the radial
colour temperature distribution, $T(R)$. The rate of energy generation per
unit area of one side of the disc, $Q$, is given by
\begin{equation}\label{eq:Q}
Q(R)= -{{\dot{M}} \over {4 \pi}} \omega R {{{\rm d} \omega} \over {{\rm d}
R}} \left[ 1 - {{\omega(R_{\rm ms}) R_{\rm ms}^2} \over {\omega R^2}} \right],
\end{equation}
where $\omega$ is the rotation velocity. The colour temperature is given by,
\begin{equation}
\label{eq:temperature}
q Q(R)= \sigma \left(T\over f_{\rm col}\right)^4,
\end{equation}
where $\sigma$ is the Stefan-Boltzmann constant and $q$ is the ratio of the
disc flux to $Q(R)$, see equation (\ref{eq:q}) below. When a fraction ($f$) of $Q(R)$
is dissipated in a corona above the disc surface (SZ94), $q<1$, and the disc
emission is due to both internal dissipation of $(1-f) Q(R)$ and
irradiation by the corona. We adopt the pseudo-Newtonian potential (Paczy\'nski
\& Wiita 1980),
\begin{equation}
\Phi(R) = -{{GM} \over {R-2R_{\rm g}}}.
\end{equation}
The rotation velocity is then
\begin{equation}
\omega(R) = {1 \over {R - 2R_{\rm g}}} \left({GM \over R}\right)^{1/2}.
\end{equation}
Then we find from equations (\ref{eq:temperature})-(\ref{eq:Q})
\begin{equation}
\label{eq:pn_temperature}
T(r) = T_0 \left[ {r - 2/3 \over r (r - 2)^3} \left(1 - {3^{3/2}\over 2^{1/2}}
{r - 2 \over r^{3/2} }\right) \right]^{1/4},
\end{equation}
where $r \equiv R/R_{\rm g}$ and
\begin{equation}\label{eq:t0}
T_0 = f_{\rm col} \left( 3 \dot M q c^6 \over 8 \pi \sigma G^2 M^2
\right)^{1/4}.
\end{equation}
At $r\gg 6$, $T(r)\simeq T_0 r^{-3/4}$. The maximum $T$ is
\begin{equation}\label{eq:tmax}
T_{\rm max} = \cases{
c_0 T_0, &${r}_{\rm in} < {r}_{\rm max}$,\cr
T({r}_{\rm in}), &otherwise,\cr}
\end{equation}
where $c_0 \simeq 0.1067$ and $r_{\rm max} \approx 9.505$.

The (unscattered) flux emitted by the disc per unit photon energy, $E$, can be
then obtained from equation (\ref{eq:fnu}),
\begin{equation}
F_E= K E^3 \int_{r_{\rm in}}^{\infty} {r {\rm d} r\over \exp[E/kT(r)] -1},
\end{equation}
where
\begin{equation}
K = (1-p_{\rm sc}) {{4 \pi} \over {h^3 c^2}} \left(G M \over c^2 D\right)^2
{\cos i \over f_{\rm col}^4},
\label{eq:disc_norm}
\end{equation}
and $h$ is the Planck constant. This model, under the name {\tt diskpn}, is a
part of the {\sc xspec} packet. Its parameters are $T_{\rm max}$, $r_{\rm in}$
and the normalization, $K$.

From these parameters, $M$ can be obtained from equation (\ref{eq:disc_norm}),
and then $\dot M$ can be obtained using equations
(\ref{eq:pn_temperature})--(\ref{eq:tmax}). For a constant $K$, $M\propto D
f_{\rm col}^2 \cos^{-1/2} i$. For constant $K$ and $T_{\rm max}$, $\dot
M\propto D^2 f_{\rm col}^ 4\cos^{-1} i$. The disc luminosity is given by
\begin{equation}\label{eq:disk_luminosity}
L_{\rm s}={2\pi D^2\over \cos i} \int F_E {\rm d} E,
\end{equation}
which, for $r_{\rm in}=6$, equals
\begin{equation}\label{eq:lum_6}
L_{\rm s}=(1-p_{\rm sc}) q {1\over 16} \dot{M} c^2={\pi\sigma G^2 M^2 T_{\rm
max}^4 \over 6 c_0^4 f_{\rm col}^4 c^4 }.
\end{equation}

\begin{figure}
\begin{center}
\leavevmode
\epsfxsize=8.4cm \epsfbox{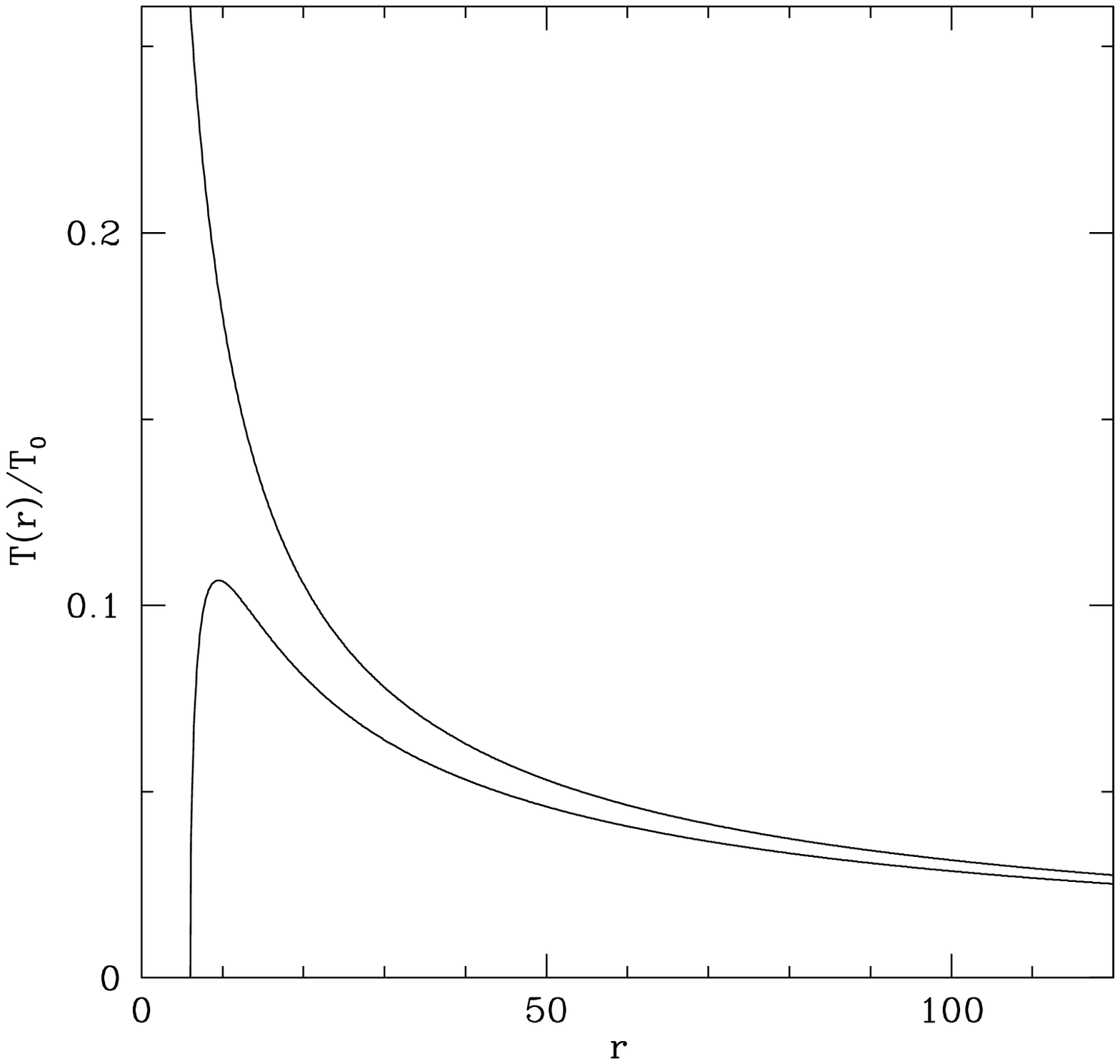}
\end{center}
\caption{The radial temperature distribution in a pseudo-Newtonian disc,
equation (\ref{eq:pn_temperature}) (lower curve) compared to the asymptotic
distribution, $\propto r^{-3/4}$, used by the {\tt diskbb} model (upper
curve).}
\label{fig:disc_temp}
\end{figure}

Finally, we point out that the above model differs significantly from a
commonly-used disc blackbody model, {\tt diskbb} (Mitsuda et al.\ 1984) in {\sc
xspec}, in which the torque-free boundary condition is neglected and the
asymptotic, $r\gg 6$, distribution is assumed, i.e., $T(r) = T_0 r^{-3/4}$.
This is a sufficiently accurate model for $r_{\rm in}\gg 6$. However, when the
disc extends close to the minimum stable orbit, the discrepancy between this
$T(r)$ and the $T(r)$ of equation (\ref{eq:pn_temperature}) becomes very
significant, as shown on Figure \ref{fig:disc_temp}. Still, a {\tt diskbb}
spectrum can be made relatively accurate with some rescaling of $R_{\rm in}$
and $T_{\rm max}$. We find in the case of $r_{\rm in}=6$ that a PN spectrum at
some $R_{\rm in}$ and $T_{\rm max}$ is close to the {\tt diskbb} spectrum
computed for $R_{\rm in}^{\tt diskbb} \approx 2.73 R_{\rm in}$ and $T_{\rm
max}^{\tt diskbb} \approx 1.04 T_{\rm max}$.

A more serious discrepancy regards the accretion efficiency. The efficiency
($\eta\equiv L/\dot M c^2$) of a PN disc extending to the minimum stable orbit
is $\eta_{\rm PN} = 1/16$, very close to the exact value in the Schwarzschild
metric, $\eta_{\rm S} = 1 - (8/9)^{1/2} \approx 0.0572$ (e.g.\ Kato, Fukue \&
Mineshige 1998), whereas the {\tt diskbb} model yields $\eta_{\tt diskbb} =
1/4$. Therefore, $\dot M$ for discs extending to the minimum stable orbit
computed using the {\tt diskbb} model and assuming $\eta_{\rm S}$ (as sometimes
done) is 4.37 times less then the actual $\dot M$.

\section{Energy balance in disc-corona systems}
\label{app:b}

We express here energy balance between a corona and an underlying disc (see 
Figure \ref{fig:geometry}) in terms of the {\it observed\/} soft, hard and 
reflected fluxes, $F_{\rm s}$, $F_{\rm h}$ and $F_{\rm r}$, respectively. This 
allows a direct comparison with results of data fitting, in which we can 
distinguish between the flux in unscattered disc photons ($F_{\rm s}$), the 
flux in scattered photons ($F_{\rm h}$), and the flux in the unscattered 
Compton-reflection and Fe K$\alpha$ photons ($F_{\rm r}$).

We first note that $F_{\rm s}$ and $F_{\rm h}$ are related in different manners
to the corresponding luminosities. The disc is optically thick and thus it
radiates with a semi-isotropic specific intensity, which implies $L_{\rm
s}=(2\pi/\cos i) D^2 F_{\rm s}$. In general, the plasma emission can be
anisotropic as well. If the emission radiated by the corona in directions away
from the disc is isotropic (which we assume in this work), $L_{\rm h} = 4\pi
D^2 F_{\rm h}$.

We consider then a corona dissipating a fraction, $f$, of the total available
power, with the underlying disc dissipating the remaining $(1-f)$ fraction
(SZ94). In general, the corona can be either homogeneous or patchy. In the
latter case, $L_{\rm s}$ includes also emission of the part of the disc not
covered by the corona. We will denote the powers dissipated in the corona and
in the disc by $P_{\rm c}$ ($=f L$) and $P_{\rm d}$ [$=(1-f)L$], respectively.
From energy conservation,
\begin{equation}
L=P_{\rm c}+ P_{\rm d}= L_{\rm h}+L_{\rm s}+L_{\rm r}.
\end{equation}
We will allow here for an anisotropic corona emission, in which the power
emitted down equals $d$ times the power emitted up. Also, we will include the
power of Compton-reflected photons scattered in the corona in the total
emission of the corona, $(1+d) L_{\rm h}$. Thus, the total emission of the
corona includes its own dissipation and the power in photons emitted by the
disc and then scattered,
\begin{equation}
(1+d)L_{\rm h} = P_{\rm c}+p_{\rm sc} (P_{\rm d}+ d L_{\rm h}),
\end{equation}
where $p_{\rm sc}$ is the average probability of scattering of a photon emitted
by the disc. This can be solved for
\begin{equation}
L_{\rm h}={P_{\rm c}+p_{\rm sc} P_{\rm d}\over 1+ d(1-p_{\rm sc}) }.
\end{equation}
The luminosity in unscattered soft photons emitted by the disc includes parts
due to the intrinsic disc emission and due to the reemission of the incident
coronal emission,
\begin{equation}
L_{\rm s} = (1-p_{\rm sc})\left[P_{\rm d}+d (1-a) L_{\rm h}\right]
\end{equation}
(where $a$ is the integrated albedo of the disc). The above equations can be
solved for
\begin{equation}\label{eq:ls}
L_{\rm s} = {(1-p_{\rm sc})\left\{ \left[1+d(1-a p_{\rm sc})\right] P_{\rm d} +
d (1-a) P_{\rm c}\right\} \over 1+d(1-p_{\rm sc})} .
\end{equation}
The ratio of the soft flux at the disc surface to the total dissipation rate per unit area is
\begin{equation}\label{eq:q}
q= {L_{\rm s}\over (1-p_{\rm sc}) L}
\end{equation}
(see Appendix A). Finally, the luminosity in unscattered Compton-reflected
photons is
\begin{equation}
L_{\rm r} = (1-p_{\rm sc}) d a L_{\rm h}.
\end{equation}
The solution for $f$ is
\begin{equation}\label{eq:f}
f={1+d(1-a p_{\rm sc})- p_{\rm sc}(1-p_{\rm sc})^{-1} L_{\rm s}/L_{\rm h}\over
1 +d a(1-p_{\rm sc}) + L_{\rm s}/L_{\rm h}}.
\end{equation}

We note that our results are consistent with those of Poutanen \& Svensson
(1996). However, their condition of energy balance is expressed in terms of the
total soft emission of the disc (denoted by $l_{\rm s}$), which includes the
part scattered in the corona, and the total emission of the corona ($l_{\rm
c}$), which includes the (unobserved) coronal emission directed towards the
disc surface. For comparison with observations, those quantities need be
transformed to the luminosities corresponding to observed fluxes; e.g., $l_{\rm
c}$ corresponds to $(1+d)$ times the actual luminosity of the corona.

\bsp
\label{lastpage}

\end{document}